\documentclass[manuscript,screen]{acmart}

\usepackage[linesnumbered,ruled,vlined]{algorithm2e} % For algorithm environment
\usepackage{amsmath} % For mathematical equations and expressions
\usepackage{amsthm} % For theorem-like environments (optional)
\usepackage{graphicx} % For including graphics or images if needed
\usepackage{float} % For controlling figure placement
\usepackage{multirow} % For tables with multi-row cells (optional)
\usepackage{tabularx} % For advanced table formatting
\usepackage{xcolor} % For colored text or highlights
\usepackage{hyperref} % For hyperlinks in the document
\usepackage{dsfont}
\usepackage{tcolorbox}
\usepackage{enumitem}
\SetKwInput{KwIn}{Input} % Customize 'Input' keyword
\SetKwInput{KwOut}{Output} % Customize 'Output' keyword
\SetKwInput{KwParam}{Parameter}
\usepackage{pifont}

\definecolor{LightRed}{rgb}{1,0.3,0.3}
\usepackage[table]{xcolor}
\newcommand{\best}[1]{\cellcolor{gray!20}\textbf{#1}}
\usetikzlibrary{patterns}
\usepackage{xcolor}
\usepackage[normalem]{ulem} % for \sout (strikeout)

\newcommand{\added}[1]{\textcolor{red}{#1}}

\usepackage{amssymb}

\usepackage{tikz}
\usepackage{pgfplots}
\usetikzlibrary{patterns}
\pgfplotsset{compat=1.18}

\usepackage{pgfplots}
\pgfplotsset{compat=1.18}
\usetikzlibrary{patterns}
\AtBeginDocument{%
  }

\begin{document}

\title{Led to Mislead: Adversarial Content Injection for Attacks on Neural Ranking Models}

\author{Amin Bigdeli}
\orcid{0009-0003-8977-9312}
\affiliation{%
  \institution{University of Waterloo}
  \city{Waterloo}
  \state{ON}
  \country{Canada}
}
\email{abigdeli@uwaterloo.ca}

\author{Amir Khosrojerdi}
\orcid{0009-0002-4341-9066}
\email{amir.khosrojerdi@mail.utoronto.ca}
\affiliation{%
  \institution{University of Toronto}
  \city{Toronto, Ontario} \country{Canada}
}

\author{Radin Hamidi Rad}
\orcid{0000-0002-9044-3723}
\email{radin.hamidi-rad@mila.quebec}
\affiliation{%
 \institution{Mila – Quebec AI Institute}
 \city{Montreal}
 \state{QC}
 \country{Canada}
 % \streetaddress{}
}

\author{Morteza Zihayat}
\orcid{0000-0002-1144-7364}
\email{mzihayat@torontomu.ca}
\affiliation{%
  \institution{Toronto Metropolitan University}
  \city{Toronto, Ontario} \country{Canada}
}

\author{Charles L. A. Clarke}
\orcid{0000-0001-8178-9194}
\affiliation{%
  \institution{University of Waterloo}
  \city{Waterloo}
  \state{ON}
  \country{Canada}
}
\email{claclark@gmail.com}

\author{Ebrahim Bagheri}
\orcid{0000-0002-5148-6237}
\affiliation{%
  \institution{University of Toronto}
  \city{Toronto}
  \state{ON}
  \country{Canada}
}
\email{ebrahim.bagheri@utoronto.ca}

\renewcommand{\shortauthors}{Amin Bigdeli et al.}
\begin{abstract}
Neural Ranking Models (NRMs) are central to modern information retrieval but remain highly vulnerable to adversarial manipulation. Existing attacks often rely on heuristics or surrogate models, limiting effectiveness and transferability. We propose \texttt{CRAFT}, a supervised framework for black-box adversarial rank attacks powered by large language models (LLMs). \texttt{CRAFT} operates in three stages: adversarial dataset generation via retrieval-augmented generation and self-refinement, supervised fine-tuning on curated adversarial examples, and preference-guided optimization to align generations with rank-promotion objectives. Extensive experiments on the MS~MARCO passage dataset, TREC Deep Learning 2019, and TREC Deep Learning 2020 benchmarks show that \texttt{CRAFT} significantly outperforms state-of-the-art baselines, achieving higher promotion rates and rank boosts while preserving fluency and semantic fidelity. Moreover, \texttt{CRAFT} transfers effectively across diverse ranking architectures, including cross-encoder, embedding-based, and LLM-based rankers, underscoring vulnerabilities in real-world retrieval systems. This work provides a principled framework for studying adversarial threats in NRMs, underscores the risks of generative AI in rank manipulation, and provides a foundation for developing more robust retrieval systems. To support reproducibility, we publicly release our source code, trained models, and prompt templates.\footnote{\added{\url{https://github.com/aminbigdeli/CRAFT}}}
\end{abstract}





\maketitle

\section{Introduction}

Information Retrieval (IR) systems have evolved significantly over the past few years, transitioning from traditional term frequency-based methods \cite{robertson2009probabilistic} to Neural Ranking Models (NRMs) that leverage deep learning architectures to better understand semantic relevance and user intent \cite{nogueira2020document,pradeep2021expando,lin2022pretrained,mitra2017neural}. NRMs now serve as the foundation of modern search platforms and enable more accurate and context-aware retrieval aligned with users’ information needs. However, studies have shown that deep neural network–based models are highly vulnerable to adversarial attacks \cite{narodytska2017simple,fawaz2019adversarial,akhtar2018threat,ebrahimi2017hotflip}, raising concerns that similar vulnerabilities may also threaten the integrity of IR systems.

Earlier generations of retrieval models were vulnerable to traditional spam techniques \cite{castillo2011adversarial}, where attackers artificially inflated term frequencies to boost the rank of target documents \cite{castillo2011adversarial,webspam}. While effective in early systems, such attacks were often detectable through simple filters and heuristics \cite{sasaki2005spam}. In contrast, recent adversarial threats exploit the vulnerabilities of NRMs in more subtle ways. By generating fluent, contextually relevant text, adversaries can promote a target document without explicitly satisfying the query by exploiting the semantic understanding capabilities of these systems. In this context, the attacker manipulates a payload document, the adversary's target content containing commercial, political, or misinformation material, by injecting an adversarial vector, a crafted text fragment designed to boost ranking while remaining imperceptible to both human readers and automated detection systems \cite{bigdeli2024empra,idem,pat}.

Recent studies have shown that NRMs are vulnerable to adversarial attacks through techniques such as token-level substitutions~\cite{prada}, adversarial triggers~\cite{brittle,pat}, and generative sentence injections~\cite{idem,bigdeli2024empra,attchain}. This is because even subtle modifications to a target document can substantially alter its position in ranked results by NRMs~\cite{idem,bigdeli2024empra}. Such vulnerabilities compromise the reliability of IR systems and open the door to disinformation campaigns and unethical practices such as black-hat SEO, which manipulates search rankings for competitive advantage~\cite{webspam}. A central challenge in designing and evaluating adversarial attacks is preserving the adversarial intent of the perturbation. Excessive or poorly controlled modifications can lead to semantic drift, diluting the original message. In contrast, effective perturbations must remain stealthy, improving rankings while preserving the document's fluency and quality. This balance is essential to avoid raising suspicion among users or triggering automated detection systems.

However, current attack methods exhibit several important limitations. \textit{First}, they are typically unsupervised and rely on heuristic optimization without ground-truth supervision. \textit{Second}, they depend on surrogate models to guide the attack process, which introduces inconsistencies due to the imperfect alignment between the surrogate and the victim model. \textit{Third}, they lack mechanisms for fine-grained control over the quality and positioning of perturbations, making the generated attacks brittle and less transferable across ranking tasks. \textit{Finally}, some methods excessively alter the payload document, causing semantic drift that undermines the core message and compromises the attack objective.

The emergence of Large Language Models (LLMs) as powerful text generation tools has enabled significant advances across a range of tasks, particularly in Information Retrieval (IR), including synthetic document generation~\cite{arabzadeh2024adapting,bigdeli2025fsap,alaofi2024generative,askari2023test,braga2024synthetic,askari2023expand} and annotation tasks \cite{rahmani2024llm4eval,upadhyay2024umbrela,wang2024human}. At the same time, their rapid progress introduces new risks in this domain. Beyond their utility for beneficial applications, LLMs can be misused as adversarial instruments for rank manipulation, producing fluent and contextually appropriate attack vectors at scale~\cite{attchain}.
% Such misuse poses serious concerns for both technical and societal domains, including the spread of misinformation and the boost of disinformation campaigns. These risks underscore the broader challenges associated with the unintended consequences of generative AI, underscoring the urgent need for comprehensive risk assessment frameworks and robust defensive mechanisms to safeguard the integrity of IR systems.
Such misuse raises serious concerns across both technical and societal domains, including the spread of misinformation~\cite{khalid2025sentiment,mohawesh2025truth} and the boosting of disinformation campaigns~\cite{DBLP:conf/icer/MartinDZCRL25,DBLP:journals/iando/BarrettFJ25}. These risks highlight the broader challenges posed by the unintended consequences of generative AI. This emphasizes the urgent need for comprehensive risk assessment frameworks and robust defense mechanisms to preserve the integrity of IR systems.

In this paper, we investigate the risks that LLMs pose to the integrity of IR systems by examining how they can be trained and fine-tuned to perform adversarial rank attacks. We introduce \texttt{CRAFT}, \textit{\textbf{C}ontext-awa\textbf{R}e \textbf{A}dversarial \textbf{F}ine-\textbf{T}uning of LLMs for Rank Attacks}. Unlike prior unsupervised and heuristic-based approaches, \texttt{CRAFT} introduces a principled pipeline that integrates high-quality adversarial dataset generation, fine-tuning strategies, and model adaptation. Importantly, \texttt{CRAFT} operates in a black-box setting where the adversary can only issue queries to the victim model and observe ranking outputs, without requiring access to gradients or internal parameters. This end-to-end framework enables large language models (LLMs) to generate effective and stealthy adversarial perturbations in real time.

To enable supervised fine-tuning of LLMs for adversarial rank attacks, \texttt{CRAFT} constructs high-quality datasets by integrating Retrieval-Augmented Generation (RAG) with a self-refinement mechanism. For each query, a neural ranker retrieves the top-ranked documents to establish a retrieval context. Conditioned on this context, the LLM generates candidate adversarial sentences aimed at promoting a payload document. These adversarial vectors are inserted at multiple positions within the payload, creating diverse adversarial variants. The modified documents are then re-ranked by the victim model, and the observed shifts in rank serve as explicit supervision signals that link perturbations to their effectiveness. To enhance dataset quality, \texttt{CRAFT} employs a self-refinement mechanism based on a \textit{chain-of-thought} feedback loop~\cite{DBLP:conf/nips/Wei0SBIXCLZ22}. In this process, the LLM iteratively evaluates ranking outcomes and revises its adversarial generations, discarding ineffective perturbations and refining promising ones. This iterative feedback not only strengthens the attack signal but also ensures that the generated perturbations remain fluent, coherent, and difficult to detect. By combining RAG for contextualized generation with chain-of-thought refinement for iterative improvement, \texttt{CRAFT} produces a supervised dataset that captures both successful and unsuccessful adversarial behaviors, providing a robust foundation for training LLMs to generate attack vectors in real time.

Once the supervised dataset has been constructed, we fine-tune LLMs directly on this data to enable them to generate adversarial perturbations conditioned on query--document pairs. This step equips the model with the ability to reproduce perturbations that are both effective and contextually coherent, providing a strong baseline capability beyond heuristic methods. To further enhance attack quality, we also incorporate a reward-based optimization stage that leverages preference signals derived from ranking outcomes. When multiple adversarial variants are produced for the same query--document pair, their relative improvements in ranking provide implicit reward signals. The LLM is then optimized to favor perturbations that achieve stronger rank promotion while preserving stealth and fluency. Together, supervised fine-tuning and reward-based optimization enable the model to reliably generate high-quality adversarial text in real-time scenarios.

To assess the effectiveness of \texttt{CRAFT}, we conduct extensive experiments on the MS MARCO V1 passage collection~\cite{msmarco}, the TREC Deep Learning 2019~\cite{TREC2019} and 2020~\cite{TREC2020} benchmarks, widely adopted benchmarks for adversarial attack studies \cite{idem,bigdeli2024empra,prada,pat,attchain}. Our results demonstrate that \texttt{CRAFT} significantly outperforms state-of-the-art baselines across all three benchmarks, with adversarial modifications consistently promoting a large fraction of target documents into the Top-10 and Top-50 results while delivering substantial average rank boosts. The framework generalizes effectively across diverse victim neural ranking models, including cross-encoder rankers, embedding-based models, and LLM-based rerankers, demonstrating robustness against a wide range of NRM architectures. At the same time, the generated perturbations preserve fluency and semantic fidelity, remain largely imperceptible to detection mechanisms and spam filters, and achieve the best balance between attack performance and stealth among all evaluated methods.
 
More concretely, the main contributions of this paper can be enumerated as follows:
\begin{enumerate}
    \item We introduce \texttt{CRAFT}, a supervised framework for adversarial rank attacks that operates in a black-box setting, going beyond heuristic and surrogate-based approaches. 
    
    \item We design a structured dataset generation pipeline that integrates retrieval-augmented generation with self-refinement, enabling the creation of high-quality adversarial training data. 
    
    \item We develop a two-stage training strategy that combines supervised fine-tuning and preference-guided optimization, allowing large language models to generate effective and covert adversarial content. 
    
    \item We conduct extensive experiments across three benchmarks and diverse victim architectures, demonstrating that \texttt{CRAFT} outperforms state-of-the-art baselines in attack success, ranking effectiveness, and linguistic imperceptibility, while generalizing across different query distributions and ranking paradigms.
\end{enumerate}

% \begin{itemize}
%     \item We introduce \texttt{CRAFT}, a supervised framework for adversarial rank attacks that goes beyond heuristic and surrogate-based methods, enabling systematic black-box exploitation of NRMs vulnerabilities using adversarial content generated by LLMs.

%     \item We design a systematic dataset construction process that integrates Retrieval-Augmented Generation (RAG) with a chain-of-thought self-refinement mechanism to produce high-quality training data for adversarial fine-tuning.
    
%     \item We fine-tune LLMs on the constructed dataset and use a reinforcement learning method with preference signals to help them generate adversarial content that is both effective and covert.

%     \item We demonstrate through extensive experiments on the MS MARCO V1 passage collection~\cite{msmarco} that \texttt{CRAFT} achieves near-perfect attack success rates, outperforms state-of-the-art baselines, and transfers robustly across diverse NRMs while preserving the integrity of the target documents.
% \end{itemize}

% The remainder of this paper is organized as follows. Section~\ref{sec:related} reviews related work on adversarial attacks in neural ranking. Section~\ref{sec:problem} formalizes the problem setting and Section~\ref{sec:methodology} presents our proposed CRAFT framework in detail. Section~\ref{sec:experiments} describes the experimental setup, and Section~\ref{sec:results} reports and analyzes the results. Finally, Section~\ref{sec:conclusion} concludes with a summary of findings and directions for future work.

\section{Related Work}
\label{sec:related}
Adversarial attacks in information retrieval have been studied under two primary objectives: \textbf{(1) corpus poisoning adversarial attacks}, which focus on degrading overall retrieval effectiveness by corrupting the training or retrieval corpus in ways that mislead the ranking model; \textbf{(2) query-targeted adversarial attacks}, aim to promote a designated payload document for a given query, thereby manipulating the ranking to increase its visibility.

% \subsection{Corpus Poisoning Adversarial Attacks}

Several studies have investigated adversarial attacks aimed at degrading retrieval effectiveness by injecting meaningless or low-quality documents into the corpus \cite{li2025unsupervised,li2025reproducing,zhong2023poisoning,wang2025tricking,su2024corpus}. The goal of these approaches is to expose vulnerabilities in retrieval models by generating adversarial documents that disrupt relevance estimation and reduce overall retrieval performance and user satisfaction. Unlike query--targeted attacks, such corpus poisoning strategies can operate in a query--independent manner, introducing noise that broadly harms retrieval across the collection \cite{li2025unsupervised,su2024corpus}. For example, \citet{li2025unsupervised} developed a query-independent corpus poisoning attack for dense retrieval that operates in the embedding space rather than the lexical space. By combining a reconstruction model with a perturbation model, their approach generates uninformative yet high-ranking adversarial documents without requiring knowledge of the query distribution. \citet{su2024corpus} introduced a structured gradient-based corpus poisoning attack through selecting high quality token perturbations. Their approach achieves high success rates across datasets and retrievers, transfers to unseen queries and domains, and extends to settings such as knowledge poisoning in RAG systems.

Alternatively, corpus poisoning attacks may be designed to target clusters of semantically related queries, thereby amplifying their disruptive effect within specific topical regions of the corpus \cite{li2025reproducing,zhong2023poisoning}. \citet{zhong2023poisoning} proposed a gradient-based corpus poisoning attack that perturbs tokens to generate adversarial passages retrievable across many queries. Using clustering to target query groups, they show that even a small number of injected passages can mislead dense retrievers in- and out-of-domain, exposing significant vulnerabilities. \citet{li2025reproducing} revisited HotFlip-based corpus poisoning attacks, improving efficiency with query clustering and centroid-based optimization. They further evaluated transfer-based black-box and query-agnostic settings, showing limited cross-model transferability but strong impact when only a small fraction of adversarial passages is injected.

% \subsection{Query-Targeted Adversarial Attacks}
In contrast to corpus poisoning, query-targeted rank attacks concentrate on manipulating retrieval results for individual queries. Central to these attacks is the notion of a payload document, defined as the specific document that the adversary seeks to artificially promote within the ranked list. Such payloads may contain malicious content (e.g., misinformation or harmful material) or be commercially motivated (e.g., SEO-driven promotion of products or services), with the overarching goal of maximizing their visibility by elevating their ranking position.

Prior works have explored query-targeted rank attacks on both (1) retrieval models \cite{liu2023black,song2022trattack} and (2) neural ranking models \cite{idem,bigdeli2024empra,prada,pat,bigdeli2025fsap,attchain}. In this paper, we focus specifically on adversarial attacks against black-box NRMs, which power most modern IR systems, and study how adversarial manipulation on payload documents can achieve rank promotion while preserving fluency and imperceptibility. We review recent work on adversarial attacks on NRMs through the lens of perturbation granularity, ranging from token-level substitutions to document-level perturbations.

\noindent \textbf{Token-level perturbations.} These attacks operate by replacing individual words in the payload document with semantically similar alternatives, subtly altering lexical content while preserving fluency. \citet{prada} proposed \texttt{PRADA}, a word-level substitution attack that leverages a surrogate model via pseudo-relevance feedback to identify important tokens and then replaces them with semantically similar alternatives. These methods are effective in black-box settings but risk semantic drift when modifying payload text.

\noindent \textbf{Trigger-based perturbations.} These attacks work by inserting short adversarial token sequences (triggers) into a document, often at position-sensitive locations such as the beginning. For example, \citet{brittle} proposed \texttt{Brittle-BERT} and demonstrated that inserting a small set of adversarial triggers into documents can cause dramatic rank promotion or demotion. These triggers are often position-sensitive, with the document head being most vulnerable. Similarly, \citet{pat} introduced \texttt{PAT}, a surrogate imitation model to generate short adversarial triggers. \texttt{PAT} integrates fluency and next-sentence prediction constraints to insert the perturbations. Trigger-based methods are compact and transferable; however, the inserted adversarial phrases can still appear incongruent to attentive human readers and may be detected by automatic filtering mechanisms, as discussed in \cite{idem,bigdeli2024empra}.

\noindent \textbf{Sentence-level perturbations.} This type of attacks involve injecting entire synthetic sentences into the document that bridge the query and document while maintaining fluency and leaving the original text unaltered. For instance, \citet{idem} proposed \texttt{IDEM}, which employs a generative language model (BART) to create fluent connection sentences that are merged into the target document, balancing coherence and adversarial impact. \citet{bigdeli2024empra} introduced \texttt{EMPRA}, which perturbs sentence embeddings to produce adversarial sentences that align with the query while preserving coherence. These sentences when added to the payload documents can boost their ranking. Unlike surrogate-based methods, \texttt{EMPRA} is surrogate-agnostic, demonstrating robustness across various neural ranking models.

\noindent \textbf{Document-level perturbations.} \citet{liu2024multi} proposed \texttt{RL-MARA}, a reinforcement learning-based framework that combines perturbations across word, phrase, and sentence granularities, formulating the attack as a sequential decision-making process guided by a surrogate ranking model. While this multi-granular strategy enhances attack flexibility, it introduces additional complexity through surrogate model training and reinforcement learning optimization. We note that the implementation of \texttt{RL-MARA} is not publicly available, and the technical details provided in the paper were insufficient to support faithful reproduction. Building on this line of work, \citet{attchain} proposed \texttt{AttChain}, which leverages LLMs with chain-of-thought prompting to iteratively rewrite larger spans of text in the target document, a successor to \texttt{RL-MARA} and represents the authors' most recent and accessible contribution in this line of research. By anchoring modifications to higher-ranked competitor documents, \texttt{AttChain} incrementally transforms the payload until rank promotion is achieved. Unlike surrogate-based approaches, \texttt{AttChain} does not rely on training a substitute model and instead issues repeated queries to the target NRM. However, this reliance on repeated querying of the victim model raises practical concerns, as it assumes continuous and unrestricted access to the target ranking system, which may be costly or infeasible in real-world settings where API rate limits, query costs, or access restrictions apply. Moreover, the iterative rewriting process substantially alters the original content, raising concerns about semantic fidelity and alignment with the payload's intended stance. Such extensive modifications may inadvertently undermine the attack objective by drifting away from the targeted message.

% \added{In addition to targeted attack methods, recent work has explored using LLMs in a zero-shot prompting setting, referred to as \texttt{LLM-Prompt}~\cite{bigdeli2025fsap}, to generate adversarial documents given a target document and a search query. Rather than proposing a novel attack strategy, \texttt{LLM-Prompt} serves as a reference point for quantifying how effectively general-purpose LLM capabilities can manipulate rankings without targeted optimization, thereby contextualizing the gains achieved by specialized attack frameworks.}

Despite their effectiveness, existing attacks against NRMs exhibit several notable limitations. Token-level substitution methods such as \cite{prada} rely heavily on surrogate models, making them costly to construct and brittle when the surrogate diverges from the victim model. Trigger-based approaches like \cite{brittle} and \cite{pat} demonstrate that short token sequences can be highly transferable, yet the resulting perturbations often appear incongruent or can be detected by automatic filters, raising concerns about stealthiness. Sentence-level injection methods such as \cite{idem} and \cite{bigdeli2024empra} aim to improve imperceptibility, but \cite{idem} remains sensitive to prompt design, while embedding perturbations of \cite{bigdeli2024empra}, although surrogate-agnostic, risk semantic drift when generating adversarial sentences. Finally, document-level strategies like \cite{attchain} avoid surrogates altogether by directly querying the victim NRM, but they require extensive iterative rewriting of the payload, which undermines efficiency and may inadvertently alter the stance or meaning of the original content, leading to attack objective failure. Moreover, this reliance on continuous access to the victim model during the attack process is impractical under realistic attack conditions and access restrictions constrain the attacker's interaction with the target system.

These methods typically address isolated aspects of the problem and lack a cohesive framework that jointly optimizes attack success and content quality. Moreover, while recent work has demonstrated that LLMs can generate adversarial content through zero-shot or few-shot prompting~\cite{attchain}, such approaches lack task-specific optimization and produce outputs that vary in quality and effectiveness across queries, as the model has no learned representation of what constitutes a successful adversarial perturbation. Gradient-based surrogate methods, on the other hand, are fundamentally bottlenecked by the alignment between the surrogate and the victim model, and their effectiveness degrades when the two diverge in architecture or training data.

This work addresses the core limitations of previous adversarial ranking methods. Unlike prior approaches, our framework is guided by the actual behavior of neural rankers rather than heuristics or surrogate models. We adopt a three-stage framework that first generates supervised adversarial data grounded in actual victim NRM feedback through an offline dataset construction phase, then fine-tunes large language models on this data through supervised learning, and finally refines their generation through preference-guided optimization, enabling them to produce more natural, transferable, and effective perturbations in a single forward pass without requiring iterative querying or surrogate model access at inference time.

\section{Threat Model}

\label{sec:threat_model}

In this section, we specify the threat model governing the adversarial setting by defining the attack objectives, the attacker's background knowledge, and the attacker's capabilities.

\subsection{Attack Objectives}
\label{sec:problem}

Consider a set of $n$ user queries \( \mathcal{Q} = \{q_1, q_2, \dots, q_n\} \) submitted to an IR system. For every query $q \in \mathcal{Q}$, the target victim neural ranking model generates an ordered list 
of candidate documents 
$\mathcal{D}_q = \{d_1, d_2, \dots, d_m\}$. The ordering of documents 
is governed by a relevance scoring function 
$f_{\mathrm{rel}} : \mathcal{Q} \times \mathcal{D} \rightarrow \mathbb{R}$,
which maps each query-document pair to a real-valued score. 
These scores quantify the predicted degree of relevance between 
the query and the document, and the ranking is obtained by sorting 
documents in $\mathcal{D}_q$ in descending order of their corresponding 
$f_{\mathrm{rel}}(q, d)$ values, such that:

\begin{equation}
    \text{Rank}(q, d) = \sum_{d' \in \mathcal{D}_q} \mathds{1}\left[ f_{\text{rel}}(q, d') > f_{\text{rel}}(q, d) \right]
\end{equation}

An adversarial attack is characterized by two fundamental components: (1) \textbf{Attack Payload}, the target document that the adversary aims to promote in the ranked list. The payload represents the content intended to be delivered to the user once the manipulation succeeds. The objective of the attack is to embed an \emph{attack vector} within this payload while preserving a high degree of \emph{fidelity}, meaning it should remain unchanged or undergo only minor perturbations that do not compromise the core malicious intent it is designed to convey; and (2) \textbf{Attack Vector}, the adversarial text injected into the payload. The vector provides indirect relevance signals that exploit the vulnerabilities of the ranking model. It must be linguistically fluent, semantically coherent, and stealthy enough to evade detection by readers and automated spam filters, while effectively boosting the rank of the payload.

Given a query $q$, a target document $d_t \in \mathcal{D}_q$ that serves as the attack payload, and a rank threshold $k \in \mathbb{N}$, the adversarial objective is to construct a perturbed version of the target document, expressed as 
$d^{\text{adv}}_t = d_t \oplus v$, where $\oplus$ denotes the injection of the attack vector $v$ into the payload, such that: 

\begin{equation}
\label{eq:rank}
    \text{Rank}(q, d^{\text{adv}}_t) \leq k,
\end{equation}

The adversarial attack is subject to the following semantic–linguistic objectives:

\begin{enumerate}[label=\textit{O\arabic*}]
\label{list:objectives}
    \item \textbf{Content Fidelity.} The adversarial document $d^{\text{adv}}_t$ should preserve the semantic coherence and key information of the original payload $d_t$, ensuring that its core intent remains unchanged.
    \item \textbf{Linguistic Fluency.} The injected attack vector $v$ should be grammatically correct, fluent, and contextually consistent, while remaining difficult to detect by automated filters.
    \item \textbf{Stealth and Effectiveness.} The perturbed document should appear natural with respect to the target query while successfully elevating the payload into the top-$k$ ranked results.
\end{enumerate}

Therefore, the task is to find a transformation function 
$\phi : \mathcal{D} \times \mathcal{Q} \rightarrow \mathcal{D}^{\text{adv}}$ 
that seeks to meet the ranking objective in Equation~\ref{eq:rank} and the mentioned three semantic–linguistic objectives ($O_{[1-3]}$) with respect to the victim NRM $f_{\text{rel}}$.

\begin{figure*}[t]
\centering 
\includegraphics[width=0.9\textwidth]{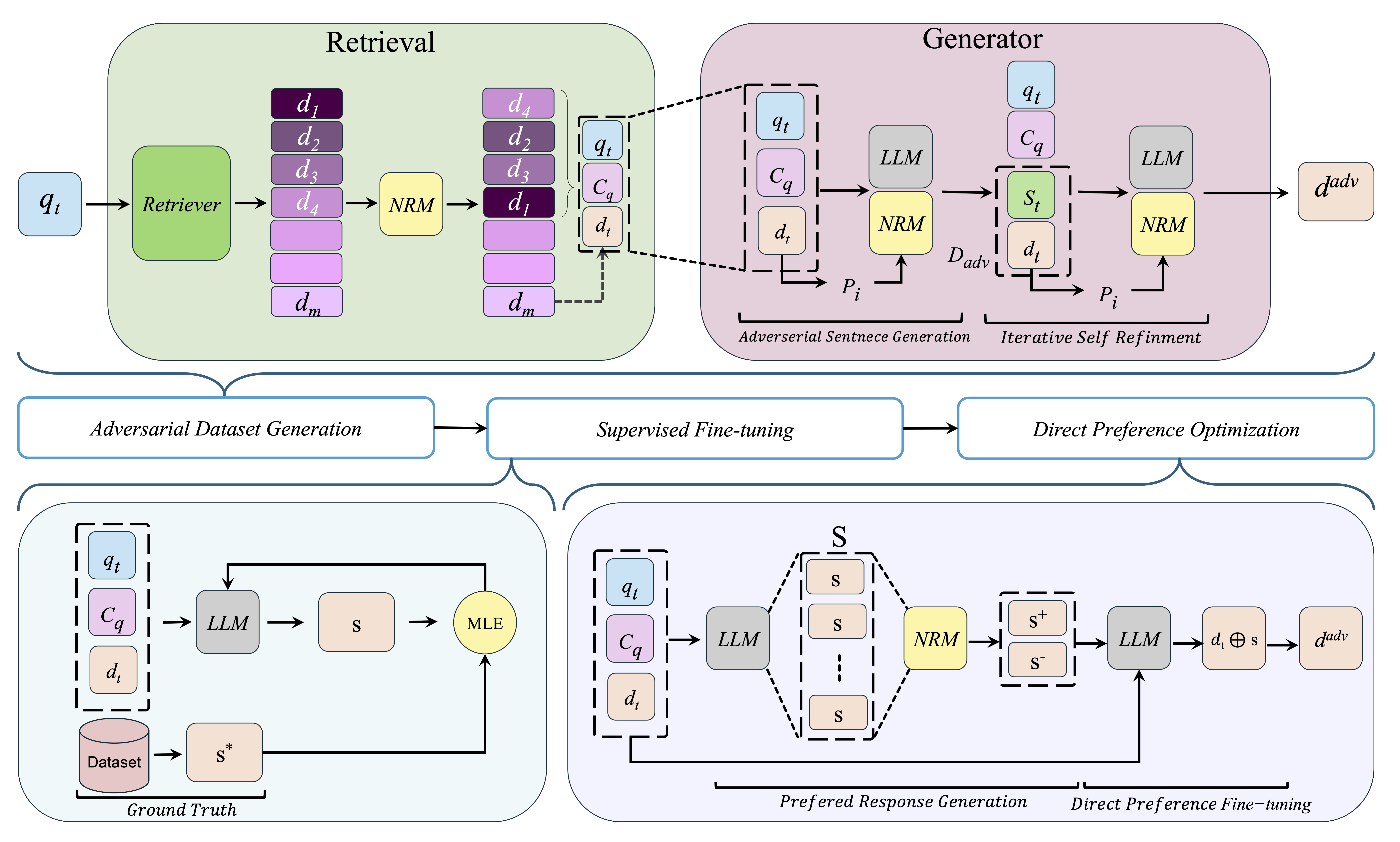} 
\caption{Overview of the \texttt{CRAFT} framework, which consists of three stages: (a) adversarial dataset generation, (b) supervised fine-tuning, and (c) direct preference optimization. The dataset produced in Stage~1 is used for supervised fine-tuning in Stage~2, after which the model is further optimized in Stage~3 via querying the neural ranking model as the policy model.
}
\label{fig:workflow_figure}
\end{figure*}

\subsection{Attacker's Background Knowledge}
\label{sec:attacker_knowledge}
 
Consistent with prior work on adversarial attacks against neural ranking models~\cite{prada,brittle,idem,bigdeli2024empra,pat,attchain}, the adversary operates under a black-box assumption where no information about the victim NRM is available, including its architecture, parameters, gradients, or training data. The attacker can only interact with the victim model by submitting queries and observing the resulting ranked list. Importantly, only the final document ordering is accessible, not the underlying relevance scores produced by $f_{\text{rel}}$. This rank-based access model reflects practical deployment conditions where commercial search systems and ranking APIs return ordered results without exposing internal scoring mechanisms.
 
\texttt{CRAFT} interacts with the victim NRM solely through query submission and rank observation, without any assumptions about the victim's architecture or internal design. This interaction occurs exclusively during the offline training phase, where the framework queries the victim NRM to collect ranking feedback for constructing the adversarial training dataset and to derive preference signals for model optimization, as detailed in Section~\ref{sec:methodology}. These queries are issued once during model preparation and are not required at inference time. Once trained, the fine-tuned model generates adversarial content conditioned on query-document pairs independently of the victim NRM, requiring no further access to the target system. The transferability of the learned adversarial patterns across multiple victim NRMs with diverse architectures is evaluated in Section~\ref{sec:cross_nrm_attack_performance_comparison}, confirming that \texttt{CRAFT} does not depend on knowledge of any specific victim architecture.

\subsection{Attacker's Capabilities}
\label{sec:attacker_capabilities}
 
The adversary controls the content of the target document $d_t$ (the attack payload) and can modify it prior to indexing by the retrieval system. This assumption is consistent with realistic attack scenarios such as web content manipulation, where an adversary can edit web pages to influence their ranking, or SEO-driven promotion of specific content. The attacker does not require any special privileges beyond query access to the victim NRM and the ability to modify the payload document. Specifically, the attacker has no ability to alter the ranking model, manipulate the corpus beyond the payload, or influence the query distribution.
 
The adversarial modification is constrained to the injection of adversarial text (the attack vector $v$) into the target document $d_t$, while the original content of the payload remains entirely unaltered, thereby preserving its core message and intended meaning. This injection-only design inherently bounds the perturbation budget by restricting modifications to appended content rather than altering existing text. Furthermore, the injected text must not semantically drift the document from its original version, preserving the payload's communicative intent while remaining undetectable by automated detection filters. These constraints, together with the semantic--linguistic objectives $O_{[1\text{-}3]}$ defined in Section~\ref{sec:problem}, collectively govern the perturbation space and ensure that adversarial modifications remain imperceptible to both human readers and automated detection mechanisms.

\section{Methodology}
\label{sec:methodology}

To address the objectives defined in Section~\ref{sec:problem}, we introduce \texttt{CRAFT}, a framework composed of three interdependent stages: (I) adversarial dataset generation, (II) supervised fine-tuning of a large language model (LLM), and (III) direct preference optimization of the LLM. \texttt{CRAFT} is designed to generate high-quality adversarial content for rank manipulation in real time, without reliance on manual tuning or heuristic sampling. The threshold criteria in Equation~\ref{eq:rank} and the objectives $O_{[1-3]}$ are enforced during dataset generation in Stage~1, and further reinforced through supervised training and optimization in Stages~2 and~3. The overall workflow of the proposed adversarial attack framework, \texttt{CRAFT}, is illustrated in Figure~\ref{fig:workflow_figure}. The figure provides a high-level overview of the three-stage pipeline and summarizes the end-to-end process. The remainder of this section details the methodology for each stage of the framework.

\subsection{Adversarial Dataset Generation}
\label{sec:dataset_generation}

The objective of the first stage of our framework is to generate a supervised dataset of high-quality adversarial examples. These examples consist of minimally modified documents that exhibit improved ranking for a given query while adhering to constraints on fluency, coherence, and indirect relevance. This dataset is used to train a transformation function capable of generating adversarial content at inference time. Formally, given a query \( q \in \mathcal{Q} \), a target document \( d_t \in \mathcal{D}_q \), and a set of top-\( c \) context documents \( \mathcal{C}_q = \{d_1, d_2, \dots, d_c\} \subset \mathcal{D}_q \), we define an adversarial sentence generator
\( \mathcal{G} : \mathcal{Q} \times \mathcal{D} \times \mathcal{C} \rightarrow \mathcal{S} \),
which outputs a candidate set \( \mathcal{S} = \{s_1, s_2, \dots, s_l\} \). Each sentence \( s \in \mathcal{S} \) is inserted into \( d_t \) at position \( p \in \mathcal{P}(d_t) \) (e.g., sentence boundaries), producing a perturbed document:  

\begin{equation}
    d_{t(p,s)}^{\mathrm{adv}} = \mathrm{Insert}(d_t,s,p)
\end{equation}

A perturbed document \( d_{t(p,s)}^{\text{adv}} \) is retained in the adversarial dataset \( \mathcal{D}_{\text{adv}} \) only if it satisfies three essential constraints. First, it must achieve rank improvement, expressed as \( \text{Rank}(q, d_{t(p,s)}^{\mathrm{adv}}) \leq \text{Rank}(q, d_t) \), which ensures that the adversarial insertion improves or at least maintains the position of the target document compared to its original version. Second, it must preserve indirect relevance, formalized as \( s \not\models q \), meaning that the adversarial sentence cannot directly answer the query or explicitly contain it. This prevents trivial detection by users or automated filters while still providing subtle cues that influence the ranker. Finally, the perturbed document must retain linguistic coherence, enforced through the condition  \( \psi(d_{t(p,s)}^{\mathrm{adv}}) > \tau \), where \( \psi \) denotes a coherence function and \( \tau \) a minimum coherence threshold. This requirement ensures that the generated adversarial sentence remains topically aligned with the query and the surrounding retrieval context, preserving covertness by anchoring the generated text to key terms captured from the query and top-ranked documents. This yields a training dataset
\( \mathcal{T} = \{(q_i, \mathcal{C}_{q_i}, d_t, s, p)\}_{i=1}^N \),
with each instance encoding the query, context, target document, perturbation, and insertion point.

To construct this training dataset, we design a generation–validation pipeline that combines a large language model (LLM) and a neural ranking model (NRM) in an iterative loop. The LLM \( \mathcal{G}(q, d_t, \mathcal{C}_q) \) proposes candidate adversarial sentences based on a query \( q \), a target document \( d_t \), and a set of context documents \( \mathcal{C} \). Each sentence is inserted into the target document and then evaluated by the NRM \( \mathcal{R}(q, d_{t(p,s)}^{\text{adv}}) \) to assess ranking shifts.

If a candidate yields a valid rank improvement within the $k$ threshold while preserving linguistic and semantic plausibility, it is retained. Otherwise, the process enters a refinement phase, where the LLM adjusts its generation using feedback from the ranking model. This feedback loop increases the likelihood of producing effective adversarial edits. The entire process results in a curated dataset of adversarial examples, each validated against the ranking model and filtered for quality. 

\begin{algorithm}[t]
\small
\setlength{\baselineskip}{0.9\baselineskip}
\caption{Stage~1: Adversarial Dataset Generation}
\label{alg:craft_stage1}
\KwIn{Query set $\mathcal{Q}$, Document corpus $\mathcal{D}$, Neural Ranking Model $\mathcal{R}$, LLM $\mathcal{G}$, Threshold $k$, Context size $c$, Iterations $n$, Coherence threshold $\tau$}
\KwOut{Adversarial training dataset $\mathcal{T}$}
Initialize $\mathcal{T} = \emptyset$ \;
\ForEach{$q \in \mathcal{Q}$}{
    $\mathcal{C}_q = \text{TopK}(\mathcal{R}(q,\mathcal{D}),c)$ \;
    \ForEach{$d_t \in \mathcal{D}$}{
        $\mathcal{S}_t = \mathcal{G}(q,d_t,\mathcal{C}_q)$ \;
        \ForEach{$p \in \mathcal{P}(d_t)$}{
            \For{$i=0$ to $n$}{
                \ForEach{$s \in \mathcal{S}_t$}{
                    $d_{t(p,s)}^{\text{adv}} = \text{Insert}(d_t,s,p)$ \;
                    \If{$\mathcal{R}(q,d_{t(p,s)}^{\text{adv}}) \leq k$ \textbf{and} $s \not\models q$ \textbf{and} $\psi(d_{t(p,s)}^{\text{adv}}) > \tau$}{
                        $\mathcal{T} \gets \mathcal{T} \cup \{(q,\mathcal{C}_q,d_t,s,p)\}$; \textbf{break} \;
                    }
                }
                $\mathcal{S}_t^{\text{high}} = \underset{s \in \mathcal{S}_t}{\arg\min} \mathcal{R}(q,d_{t(p,s)}^{\text{adv}})$ \;
                $\mathcal{S}_t = \mathcal{G}(q,d_t,\mathcal{S}_t^{\text{high}})$ \;
            }
        }
    }
}
\Return $\mathcal{T}$
\end{algorithm}
\paragraph{Adversarial Sentence Generation. }

Given a query \( q \), the first step in our method is to establish a contextual foundation for the sentence generation module \( \mathcal{G} \). To this end, we select the top-\( c \) ranked documents and assemble them into a context set \( \mathcal{C}_q = \{d_1, d_2, \dots, d_{c}\} \), where \( c \) denotes the number of documents incorporated. This set provides semantic grounding, enabling the LLM to generate sentences that are contextually aligned with the query \( q \). When injected into the target document \( d_t \), these sentences are designed to subtly enhance its ranking. To generate such adversarial sentences, the LLM is prompted to produce a pool of candidate adversarial sentences:  

\begin{equation}
   \mathcal{S}_t=\mathcal{G}(q,d_t,\mathcal{C}_q)
\end{equation}

Each sentence \( s \in \mathcal{S}_t \) is inserted into the target document at a position \( p \), resulting in a perturbed version defined as:  

\begin{equation}
    d_{t(p,s)}^{\text{adv}} = \text{Insert}(d_t, s, p)
\end{equation}

The perturbed document is then passed to the neural ranking model \( \mathcal{R} \), which evaluates its ranking position following the insertion of the adversarial sentence, as defined below:

\begin{equation}
    \text{Rank}(q, d_{t(p,s)}^{\text{adv}}) = \mathcal{R}(q, d_{t(p,s)}^{\text{adv}})
\end{equation}

If the insertion yields a ranking improvement, the adversarial sentence is retained; otherwise, the system proceeds into a feedback loop where subsequent refinements are generated and re-evaluated until the criteria are met.

\paragraph{Iterative Self-Refinement}
 
When none of the initially generated adversarial sentences achieves the required ranking improvement, the framework engages in an iterative refinement process that implements a feedback loop between the LLM generator and the neural ranking model. Specifically, the NRM evaluates each candidate perturbation and returns ranking outcomes that serve as reward signals to the generator. By conditioning subsequent generations on the most successful perturbations from prior iterations, the framework implicitly approximates gradient-based optimization over the black-box ranker, progressively steering the LLM toward more effective adversarial perturbations. In this stage, the neural ranking model identifies the most influential adversarial candidates by selecting those that yield the greatest improvement in rank:
 
\begin{equation}
    \mathcal{S}_t^{p, \text{high}} = \underset{s \in \mathcal{S}_t^p}{\arg \min} \, \text{rank}(q, d_{t(p,s)}^{\text{adv}})
\end{equation}
These top-ranked candidates are subsequently passed to the large language model (LLM), which leverages them as conditioning signals to generate refined adversarial variants. By feeding back the highest-performing perturbations, the framework balances exploration of new adversarial strategies with exploitation of patterns that have already proven effective, guiding the generator toward increasingly targeted content. The refined candidates are generated as:
 
\begin{equation}
    \mathcal{S}_t^{p, \text{new}} = G(q, d_t, \mathcal{S}_t^{p, \text{high}})
\end{equation}
 
Each of the refined adversarial sentences is then inserted into \( d_t\) to generate the updated adversarial documents:
 
\begin{equation}
    d_{p,s'}^{\text{adv, new}} = \text{Insert}(d_t, s', p), \quad \forall s' \in \mathcal{S}_t^{p, \text{new}}
\end{equation}
 
The NRM then re-evaluates these updated documents to evaluate ranking impact. This refinement loop continues until one of two stopping criteria is satisfied: (i) a document achieves the target rank threshold $k$, i.e., \( \text{rank}(q, d_{p,s'}^{\text{adv, new}}) \leq k \); or (ii) the maximum number of refinement iterations, denoted by \( n \), is reached.
 
For clarity, the overall procedure of adversarial dataset generation is summarized in Algorithm~\ref{alg:craft_stage1}. This step-by-step pseudocode complements the technical description above and provides a concise overview of the flow. This algorithm highlights the iterative nature of Stage~1, where candidate perturbations are generated, evaluated, and refined until high-quality adversarial examples are obtained. We also provide the full prompt templates used in our model in Appendix~\ref{apx:prompt} for reproducibility.

\subsection{Supervised Fine-Tuning}
\label{sec:finetuning}

To achieve adversarial content generation that meets our stated objectives, we define a transformation function 
\(\phi : \mathcal{Q} \times \mathcal{D} \times \mathcal{C} \rightarrow \mathcal{S}\) 
that maps a query \(q\), a target document \(d_t\), and a context set \(\mathcal{C}_q\) to an adversarial sentence \(s\). 
When inserted into \(d_t\), the adversarial sentence should improve the rank of the resulting document while adhering to the semantic–linguistic constraints described in Section~\ref{sec:dataset_generation}.  

To learn \(\phi\), we perform supervised fine-tuning of a parameterized model \(\phi_\theta\) (e.g., a sequence-to-sequence large language model) using Maximum Likelihood Estimation (MLE). 
In this setting, each training sample is represented as \((q, \mathcal{C}_q, d_t, s^{*}) \in \mathcal{T}\), 
where \(s^{*}\) denotes the Gold adversarial sentence obtained during Stage~1 (Section~\ref{sec:dataset_generation}). 
MLE minimizes the negative log-likelihood of these target adversarial sentences as follows:

\begin{algorithm}[t]
\small
\setlength{\baselineskip}{0.9\baselineskip}
\caption{Stage~2: Supervised Fine-Tuning (SFT)}
\label{alg:craft_stage2}
\KwIn{Adversarial dataset $\mathcal{T}$}
\KwOut{SFT model parameters $\theta_{\text{SFT}}$}
Train the model by minimizing the negative log-likelihood: \\
$\theta_{\text{SFT}} = \arg\min_\theta \; \mathbb{E}_{(q,\mathcal{C}_q,d,s^*)\sim\mathcal{T}} \left[-\log \pi_\theta(s^* \mid q,\mathcal{C}_q,d)\right]$ \;
\Return $\theta_{\text{SFT}}$
\end{algorithm}

\begin{equation}
\label{eq:mle}
    \theta_{\text{MLE}} = \arg\min_\theta \; 
    \mathbb{E}_{(q, \mathcal{C}_q, d_t, s^{*}) \sim \mathcal{T}}
    \left[-\log \pi_\theta(s^{*} \mid q, \mathcal{C}_q, d_t)\right]
\end{equation}

where \(\pi_\theta\) denotes the conditional distribution defined by the model. 
During training, the model generates a candidate adversarial sentence \(s\), 
which is aligned toward the Gold supervision signal \(s^{*}\) through this objective.  

For clarity, the procedure for supervised fine-tuning is summarized in Algorithm~\ref{alg:craft_stage2}. 
This pseudocode provides a step-by-step view of how the model parameters are optimized using MLE over the curated adversarial dataset. 
By presenting the optimization process explicitly, the algorithm complements the mathematical formulation in Equation~\ref{eq:mle} and highlights the role of supervised training in grounding the model’s ability to generate effective adversarial content.

\subsection{Direct Preference Optimization}
\label{sec:dpo}

While MLE enables the model to learn the distribution of effective adversarial sentences, it does not explicitly optimize for the attack objectives. 
To address this limitation, we employ Direct Preference Optimization (DPO) \cite{rafailov2023direct}, which formulates fine-tuning as a distributional alignment problem under preference constraints. 
In this setting, adversarial generations are compared in pairs, with preferences determined by their ranking impact and constraint satisfaction. 
These preferences act as a reward signal, encouraging the model to favor outputs that yield stronger ranking improvements while preserving fluency, coherence, and indirect relevance.  

Given an input triplet \((q, \mathcal{C}_q, d_t)\), we define an adversarial sentence \(s\) as a perturbation generated for insertion into the target document \(d_t\), producing a perturbed document \(d^{\text{adv}} = d_t \oplus s\). Candidate adversarial sentences are sampled from the supervised fine-tuned model as \(\mathcal{S} = \{ s \sim \pi_{\theta_{\text{SFT}}}(\cdot \mid q,\mathcal{C}_q,d_t) \}\), and from this set \(\mathcal{S}\), we construct two contrastive outputs:

\begin{enumerate}
    \item \(s^+\): a \emph{preferred} perturbation with the best rank within the top-$k$;
    \item \(s^-\): a \emph{rejected} perturbation is the best-ranked candidate that fails to meet the threshold. Selecting the strongest negative avoids trivial contrasts with poor candidates and provides a more informative supervision signal, enabling the model to learn the subtle distinctions between successful and unsuccessful perturbations.
\end{enumerate}

The objective is to learn a distribution \(\pi_\theta\) that prefers \(s^+\) over \(s^-\), in accordance with a latent reward model \(R(\cdot)\). 
Instead of explicitly learning \(R\), DPO updates the model by maximizing the following objective:

\begin{equation}
    \mathcal{L}_{\text{DPO}}(\theta) = - \mathbb{E}_{(q, \mathcal{C}_q, d_t)} \Bigg[ \log \sigma \Bigg( \beta \Big[ \log \pi_\theta(s^+ \mid q, \mathcal{C}_q, d_t) - \log \pi_\theta(s^- \mid q, \mathcal{C}_q, d_t) \Big] \Bigg) \Bigg]
\end{equation}

where \(\sigma(\cdot)\) is the logistic sigmoid and \(\beta > 0\) is a temperature hyperparameter controlling preference sharpness. 
The underlying reward signal is derived from whether the perturbation leads to rank improvement:  

\begin{equation}
    R(s; q, d_t, \mathcal{C}_q) = \mathbb{I}\!\left[ \text{Rank}(q, d_t \oplus s) \leq k \right]
\end{equation}

% \begin{algorithm}[t]
% \small
% \setlength{\baselineskip}{0.9\baselineskip}
% \caption{Stage~3: Direct Preference Optimization (DPO)}
% \label{alg:craft_stage3}
% \KwIn{SFT model parameters $\theta_{\text{SFT}}$, Adversarial dataset $\mathcal{T}$, Neural Ranking Model $\mathcal{R}$, Threshold $k$}
% \KwOut{Final generator $\phi_{\theta^*}$}
% \textbf{Preference Dataset Construction:} \\
% Initialize $\mathcal{D}_{\text{pref}} = \emptyset$ \;
% \ForEach{$(q,\mathcal{C}_q,d_t) \in \mathcal{T}$}{
%     $\mathcal{S}^+ = \{s : \mathcal{R}(q,d_t\oplus s) \leq k\}$ \;
%     $\mathcal{S}^- = \{s : \mathcal{R}(q,d_t\oplus s) > k\}$ \;
%     \ForEach{$s^+ \in \mathcal{S}^+$, $s^- \in \mathcal{S}^-$}{
%         $\mathcal{D}_{\text{pref}} \gets \mathcal{D}_{\text{pref}} \cup \{(q,\mathcal{C}_q,d_t,s^+,s^-)\}$ \;
%     }
% }
% \textbf{Preference Optimization:} \\
% $\theta^* = \arg\min_\theta \; \mathbb{E}_{(q,\mathcal{C}_q,d,s^+,s^-) \sim \mathcal{D}_{\text{pref}}} \left[\mathcal{L}_{\text{DPO}}(\theta)\right]$ \;
% \Return $\phi_{\theta^*}$
% \end{algorithm}

\begin{algorithm}[t]
\small
\setlength{\baselineskip}{0.9\baselineskip}
\caption{Stage~3: Direct Preference Optimization (DPO)}
\label{alg:craft_stage3}
\KwIn{SFT model parameters $\theta_{\text{SFT}}$, Adversarial dataset $\mathcal{T}$, Neural Ranking Model $\mathcal{R}$, Threshold $k$}
\KwOut{Final generator $\phi_{\theta^*}$}
\textbf{Preference Dataset Construction:} \\
Initialize $\mathcal{D}_{\text{pref}} = \emptyset$ \;
\ForEach{$(q,\mathcal{C}_q,d_t) \in \mathcal{T}$}{
     $\mathcal{S} = \{s \sim \pi_{\theta_{\text{SFT}}}(\cdot \mid q,\mathcal{C}_q,d_t)\}$ \;
     $s^{+} = \arg\min_{s \in \mathcal{S} : \mathcal{R}(q, d_t \oplus s) \leq k} \; \mathcal{R}(q, d_t \oplus s)$ \;
    
$s^{-} = \arg\min_{s \in \mathcal{S} : \mathcal{R}(q, d_t \oplus s) > k} \; \mathcal{R}(q, d_t \oplus s)$ \;

    % $\mathcal{S}^+ = \{s \in \mathcal{S} : \mathcal{R}(q,d_t\oplus s) \leq k\}$ \;
    % $\mathcal{S}^- = \{s \in \mathcal{S} : \mathcal{R}(q,d_t\oplus s) > k\}$ \;
    % \ForEach{$s^+ \in \mathcal{S}^+$, $s^- \in \mathcal{S}^-$}{
    %     $\mathcal{D}_{\text{pref}} \gets \mathcal{D}_{\text{pref}} \cup \{(q,\mathcal{C}_q,d_t,s^+,s^-)\}$ \;
    % }
}
\textbf{Preference Optimization:} \\
$\theta^* = \arg\min_\theta \; \mathbb{E}_{(q,\mathcal{C}_q,d_t,s^+,s^-) \sim \mathcal{D}_{\text{pref}}} \left[\mathcal{L}_{\text{DPO}}(\theta)\right]$ \;
\Return $\phi_{\theta^*}$
\end{algorithm}

where \(\mathbb{I}[\cdot]\) is the indicator function.  

From a probabilistic viewpoint, this amounts to aligning the learned distribution \(\pi_\theta\) with an implicit preference distribution \(\pi^*\) defined by:

\begin{equation}
\pi^*(s^+ \mid q, \mathcal{C}_q, d_t) \; \gg \; \pi^*(s^- \mid q, \mathcal{C}_q, d_t).
\end{equation}

To achieve this preference alignment, DPO minimizes the reverse-KL divergence between \(\pi_\theta\) and \(\pi^*\), weighted by preference likelihood ratios. 
This interpretation highlights DPO as a principled framework for policy refinement under pairwise preference feedback, without requiring explicit reward regression.

The preference-guided optimization stage is detailed in Algorithm~\ref{alg:craft_stage3}. 
This algorithm illustrates the construction of preference pairs from adversarial generations and their subsequent use in the DPO objective. 
By contrasting successful and unsuccessful perturbations, the algorithm shows how pairwise preferences are translated into optimization signals that refine the model’s alignment with the attack objectives.

\vspace{-1.0em}
\subsection{Inference}
\label{sec:inference}

At inference time, the optimized model \(\pi_{\theta^*}\) is deployed to generate real-time adversarial sentences \(s_{\text{adv}}\) conditioned on unseen query–document pairs \((q, \mathcal{C}_q, d_t)\). 
The process begins by retrieving a set of top-$c$ context documents for each query using the neural ranking model \(\mathcal{R}\), which provides the generator with relevant background signals to guide the perturbation process.  

For each candidate payload document, the generator samples multiple adversarial sentence candidates that are inserted at different positions within the document, yielding perturbed variants of the form \(d^{\text{adv}} = d_t \oplus s\). 
Each perturbed document is then re-scored by the ranking model \(\mathcal{R}\), and the variant that achieves the greatest rank improvement relative to the original is selected as the adversarial output.  

\begin{algorithm}[t]
\small
\setlength{\baselineskip}{0.9\baselineskip}
\caption{Adversarial Document Generation During Inference}
\label{alg:craft_inference}
\KwIn{Test queries $\mathcal{Q}_{\text{test}}$, Corpus $\mathcal{D}$, Neural Ranking Model $\mathcal{R}$, Fine-tuned generator $\phi_{\theta^*}$, Context size $c$}
\KwOut{Generated adversarial documents $\mathcal{D}_{\text{adv}}$}
Initialize $\mathcal{D}_{\text{adv}} = \emptyset$ \;
\ForEach{$(q,d_t) \in \mathcal{Q}_{\text{test}}$}{
    $\mathcal{C}_q = \text{TopK}(\mathcal{R}(q,\mathcal{D}),c)$ \;
    $d_t^{\text{adv}} = d_t$; $\text{best\_rank} = \mathcal{R}(q,d_t)$ \;
    \ForEach{$p \in \mathcal{P}(d_t)$}{
        $\mathcal{S}_{\text{cand}} = \text{Sample}(\phi_{\theta^*},q,\mathcal{C}_q,d_t)$ \;
        \ForEach{$s \in \mathcal{S}_{\text{cand}}$}{
            $d_{\text{temp}} = \text{Insert}(d_t,s,p)$ \;
            $\text{rank}_{\text{temp}} = \mathcal{R}(q,d_{\text{temp}})$ \;
            \If{$\text{rank}_{\text{temp}} < \text{best\_rank}$}{
                $d_t^{\text{adv}} = d_{\text{temp}}$; $\text{best\_rank} = \text{rank}_{\text{temp}}$ \;
            }
        }
    }
    $\mathcal{D}_{\text{adv}} \gets \mathcal{D}_{\text{adv}} \cup \{d_t^{\text{adv}}\}$ \;
}
\Return $\mathcal{D}_{\text{adv}}$
\end{algorithm}

This inference procedure enables \texttt{CRAFT} to carry out black-box adversarial attacks in real-time, directly optimizing for ranking effectiveness without relying on iterative search or heuristic constraints. 
The complete workflow is summarized in Algorithm~\ref{alg:craft_inference}.

\section{Experimental Setup}
\label{sec:experiments}

In this section, we describe the experimental setup used to evaluate \texttt{CRAFT}. We first present the datasets, including our generated dataset in Section~\ref{sec:dataset_generation}, followed by details of the baseline methods and evaluation metrics. We then outline the implementation settings to ensure reproducibility and fair comparison. 

\noindent\textbf{Reproducibility.} To facilitate reproducibility, we provide full access to all prompts, datasets, and source code in a publicly available GitHub repository\footnote{\href{https://github.com/aminbigdeli/CRAFT}{https://github.com/aminbigdeli/CRAFT}}
. In addition, the fine-tuned models are released on HuggingFace\footnote{\href{https://huggingface.co/collections/radinrad/craft-68b50627b2b18994d0c4d701}{https://huggingface.co/collections/radinrad/craft-68b50627b2b18994d0c4d701}}
 to facilitate direct reuse and further experimentation.

\subsection{Dataset Overview}
\subsubsection{Benchmark Dataset}

Consistent with prior studies~\cite{prada,brittle,pat,bigdeli2024empra,idem}, we conduct experiments on the widely used MS MARCO passage dataset~\cite{msmarco}, encompassing 8.8 million passages treated as documents, over 500,000 training queries, and 6,980 development queries. This large-scale, well-annotated benchmark enables direct comparison with state-of-the-art methods and ensures the practical relevance of our findings for advancing document ranking methodologies.

Following prior studies~\cite{attchain,liu2024multi,bigdeli2024empra}, we additionally evaluate on the TREC Deep Learning Track 2019~\cite{TREC2019} and 2020~\cite{TREC2020} benchmarks to assess the generalizability of \texttt{CRAFT}. Each benchmark comprises 200 queries and they span a broad range of real web information needs. As a result, they provide stronger evidence that the adversarial patterns learned by \texttt{CRAFT} transfer to various unseen queries.

\subsubsection{Target Queries and Documents}
\label{sec:target_documents}
Following Wu et al.~\cite{idem,bigdeli2024empra}, we randomly sample a subset of queries from the MS MARCO development set, hereafter denoted as MS MARCO Dev. To build our adversarial datasets, we first select a pool of 1,000 target queries; the train--test partition is defined after dataset filtering and is detailed in Section~\ref{sec:dataset_variations}. For the TREC DL 2019 and TREC DL 2020 benchmarks, we use the full set of 200 queries from each track. For each query, we target two distinct types of documents, Easy-5 and Hard-5, selected from the re-ranked results produced by the primary victim neural ranking model after applying it to the top-1K BM25 retrieved documents. This dual-target approach allows us to systematically assess the impact of our rank boosting techniques on documents with varying levels of initial ranking position.

\noindent\textbf{Easy-5}: This group consists of five documents initially ranked between positions 51 and 100 in the search results. Specifically, one document is randomly sampled from every ten-ranked positions within this range (e.g., ranks 51, 63, 76, 84, and 91). By targeting these mid-ranked documents, we aim to evaluate how our augmentation strategies enhance the visibility of documents that are neither highly ranked nor too obscure.
 
\noindent \textbf{Hard-5}: In contrast, this group comprises the five lowest-ranked documents from the re-ranked list, specifically those at positions 996–1000, representing the most challenging cases for rank boosting. By focusing on these least visible documents, we critically examine the robustness of our augmentation approach when applied to content with minimal initial exposure.
 
\noindent\textbf{Mixture}: Following prior work~\cite{attchain,bigdeli2024empra,idem,prada}, we construct a limited evaluation set of \textit{Mixture} target documents from the held-out test queries of MS MARCO Dev, consisting of 100 query–document pairs sampled equally from the Easy-5 and Hard-5 categories, to provide a computationally efficient benchmark for costly evaluations.
 
\subsubsection{Generated Dataset Variations}
\label{sec:dataset_variations}
To capture a broad spectrum of adversarial effectiveness, we construct multiple dataset variants from the full pool of 1,000 randomly sampled queries from MS MARCO Dev and their corresponding target documents following the approach described in Section~\ref{sec:dataset_generation}. Each variant applies progressively stricter criteria for selecting adversarial examples, thereby reflecting different levels of rank improvement.
 
\noindent\textbf{Gold Dataset.} For every query–document pair, we retain the best-performing adversarial sentence among all insertion positions. This dataset captures the strongest augmentation instances, isolating the upper bound of rank improvement achievable through adversarial injection.
 
\noindent\textbf{Diamond Dataset.} This stricter variant retains only instances where adversarial injection yields substantial rank gains, specifically moving Easy-5 into top-10 and Hard-5 documents into the top-50. By applying this threshold, the Diamond dataset emphasizes highly impactful augmentations and provides the most reliable training signals. Accordingly, we adopt the Diamond dataset as the supervised resource for fine-tuning large language models, since it offers the clearest and most consistent examples of effective adversarial edits. As a result of applying these stricter filtering criteria, the Diamond dataset contains adversarial examples derived from 875 of the original 1,000 queries. From these, we randomly select 800 queries for training (supervised fine-tuning and preference-guided optimization). The remaining 200 queries, comprising 75 Diamond queries not used for training and 125 queries that did not pass the Diamond filtering criteria, form the held-out test set for evaluation on MS MARCO Dev.

\subsection{Target NRMs}

To evaluate the effectiveness and robustness of adversarial rank attacks, we conduct experiments against multiple victim neural ranking models. This setup compares our proposed method with the baselines under diverse set of neural ranking architectures, providing a comprehensive measure of its attack generalizability. 

We adopt \texttt{msmarco-MiniLM-L-12-v2}~\cite{sbert} as our primary victim NRM. This model is a cross-encoder fine-tuned on MS MARCO that uses the MiniLM architecture. The model has demonstrated strong retrieval effectiveness, making it a widely used benchmark in adversarial IR studies \cite{idem,bigdeli2024empra,pat}.

To further assess transferability across different neural ranking architectures, we include four additional victim models. The first two, ms-marco-electra-base\cite{sbert} and distilroberta-base\cite{nogueira2020document}, are cross-encoder rankers fine-tuned on the MS MARCO training set. We also evaluate against Qwen3-Embedding-0.6B~\cite{qwen3embedding}, a state-of-the-art embedding model that ranks among the top-10 on the MTEB benchmark~\cite{muennighoff2022mteb}, adopted here in a zero-shot reranking setting. Finally, with the growing adoption of LLM-based rerankers in modern retrieval pipelines, we investigate how adversarial attacks perform when the victim is a full-scale LLM reranker rather than a conventional encoder-based model. To this end,  we include RankZephyr~\cite{pradeep2023rankzephyr}, a Mistral-7B-based open-source listwise reranker that achieves state-of-the-art performance on standard IR benchmarks, matching or surpassing proprietary models RankGPT~\citet{sun2023chatgpt}. We adopt RankZephyr using a sliding window size of 20 and a step size of 10, implemented via the RankLLM toolkit~\cite{sharifymoghaddam2025rankllm} and following the configuration recommended in the original work. Together, these models extend the evaluation by covering cross-encoder, embedding-based, and LLM-based ranking architectures, enabling a broader examination of adversarial attack robustness across distinct neural ranking paradigms.

\vspace{-1.0em}
\subsection{Baselines}
To evaluate the effectiveness of \texttt{CRAFT}, we compare its performance against a selection of state-of-the-art baseline methods including word-level \cite{prada}, trigger-level \cite{pat,brittle}, sentence-level  \cite{idem,bigdeli2024empra}, and document-level \cite{attchain} attacks.

\texttt{PRADA} \cite{prada} is a word-level attack method that identifies key terms within the document using a surrogate ranking model and replaces tokens with their nearest synonyms in an embedding space.
\texttt{Brittle-BERT}~\cite{brittle} is a trigger-based attack method that appends adversarial trigger tokens at the beginning of the document to manipulate ranking behavior.
\texttt{PAT}~\cite{pat} is another variation of trigger-based attacks that strategically inserts trigger words at the start of the document, leveraging a surrogate model to identify optimal placements.
\texttt{IDEM}~\cite{idem} is considered a sentence-level adversarial attack method that generates up to 500 connection sentences using a fine-tuned BART model. The most effective sentence, balancing fluency and relevance, is injected into the document to manipulate rankings.
\texttt{EMPRA}~\cite{bigdeli2024empra} is another sentence-level method that generates adversarial sentences by progressively traversing from the original document context toward query-relevant semantic regions within embedding space. The resulting adversarial sentences, when injected into the target document, can boost the document’s position in the ranking results.
\texttt{AttChain}~\cite{attchain} is an attack framework that generates adversarial content in five steps, where each step uses an LLM to produce candidate perturbations. The method evaluates these perturbations with the neural ranker and chains the successful ones together to gradually improve the rank of the target document.

All baselines were implemented using the official source code repositories provided by their respective authors, and their hyperparameter configurations were set in accordance with the values reported in the original publications to ensure a fair and consistent comparison.

\vspace{-1.0em}
\subsection{Evaluation Metrics}

\subsubsection{Attack Performance}
Following prior studies~\cite{idem,bigdeli2024empra,attchain,pat,prada}, we evaluate the performance of adversarial attacks using metrics that capture both their effectiveness and the degree of rank improvement.

\noindent\textbf{Attack Success Rate (ASR) (\%).} The proportion of cases where the adversarial modification yields a strict rank improvement.

\noindent\textbf{Boosted Top-10 Rate (Top-10) (\%)} The proportion of target documents that achieve a rank within the top-10 results after adversarial modification.

\noindent\textbf{Boosted Top-50 (Top-50) (\%).} The proportion of target documents that achieve a rank within the top-50 results after adversarial modification.

\noindent\textbf{Average Boosted Rank (Boost).} The average rank improvement across all queries where adversarial modification is applied. This metric captures the magnitude of rank shifts introduced by adversarial edits.

\subsubsection{Content Fidelity Metrics}
Consistent with \cite{prada}, to evaluate how well adversarial documents preserve the semantic and structural integrity of the original payload, we employ the following metrics:

\noindent\textbf{Semantic Similarity (SS):} The average BERTScore F1~\cite{zhang2019bertscore} between the original and adversarial documents, which measures token-level similarity in contextual embedding space. This captures how closely the adversarial document preserves the semantic content of the original text while allowing for adversarial modifications.

\noindent\textbf{Adversarial Token Insertion (ATI).} The average number of tokens inserted by the adversary relative to the original document length. Lower values indicate minimal augmentation, while higher values reflect heavier modification.  

\noindent\textbf{Average Distance Tokens (ADT).} The average token-level Levenshtein distance~\cite{lcvenshtcin1966binary} between the original and adversarial documents. This metric accounts for insertions, deletions, and substitutions, where lower values denote closer lexical similarity and higher values greater divergence.

\noindent\textbf{Lexical Overlap Recall (LOR).} The average ROUGE-L recall \cite{lin2004rouge} score between the original and adversarial documents, measuring the extent to which the payload is preserved without lexical changes. Higher LOR values indicate stronger preservation of the payload content.

\subsubsection{Quality and Naturalness}
To comprehensively evaluate the quality, naturalness, and linguistic acceptability of adversarially generated documents, we adopt a diverse set of metrics capturing fluency and grammaticality. This evaluation approach is consistent with prior work in adversarial attack against neural ranking models \cite{bigdeli2024empra,idem,prada,attchain,pat}. 

\noindent\textbf{Perplexity (PPL).} To evaluate text fluency, we employ a pre-trained GPT-2 model~\cite{radford2019language} to calculate the perplexity of both original and adversarial documents. Perplexity measures how well a language model predicts a sequence of tokens, with lower values indicating greater fluency and stronger alignment with natural language usage.

\noindent\textbf{Acceptability Score (AcS).} Measures perceived text quality and its acceptability using a neural language model trained on the COLA dataset \cite{warstadt2019neural} to distinguish well-formed text from syntactically or semantically irregular constructions.

\noindent\textbf{Grammar Assessment.} We use the LanguageTool (\url{https://languagetool.org/})
API to evaluate grammatical correctness and overall quality for original and adversarial documents. Three averaged metrics are reported: \textit{Change Correctness}, \textit{Change Suggestions}, and \textit{Quality}, which evaluates how well the revised text aligns with fluent and grammatically accurate writing.

\subsection{Implementation Details}
 
\noindent\textbf{Adversarial Dataset Generation.}
To generate the adversarial dataset described in Section~\ref{sec:dataset_generation}, which serves as the supervised training resource for fine-tuning LLMs, we used Qwen3-32B~\cite{qwen3}, a state-of-the-art model known for strong semantic coherence and context-aware text generation. This choice was motivated by its ability to produce adversarial modifications that preserve fluency while embedding effective perturbations.
 
For document ranking, we query the msmarco-MiniLM-L-12-v2~\cite{sbert} neural ranker. For each user query, the five highest-ranked documents were retrieved and provided as contextual evidence, a context size selected to supply the generator with sufficient retrieval signal while remaining within practical prompt length constraints. The refinement process terminates under two conditions: (i) the adversarially modified document achieves rank $\leq k = 10$, consistent with the top-10 promotion threshold adopted in prior work~\cite{idem,bigdeli2024empra,attchain}, or (ii) the maximum of $n = 5$ iterations is reached, a value determined empirically as sufficient for convergence in the majority of cases without incurring excessive generation costs. The coherence constraint $\psi$ requires each candidate adversarial sentence to incorporate at least $\tau = 5$ query-associated key terms, or their semantically similar equivalents, extracted from the top-ranked documents and the query itself. Candidates falling below this threshold are discarded. The complete set of prompts used for both initial generation and iterative refinement is provided in Appendix~\ref{apx:prompt} and is also publicly available in our \href{https://github.com/aminbigdeli/CRAFT}{GitHub repository}.
 
\noindent\textbf{Supervised Fine-Tuning.}
For the supervised fine-tuning stage described in Section~\ref{sec:finetuning}, we adopted two state-of-the-art LLMs as transformation functions $\phi$: DeepSeek-R1-Distill-Llama-70B~\cite{deepseekr1} and QwQ-32B~\cite{qwq32b}. We denote the resulting models as \texttt{CRAFT$_{\text{Llama3.3}}$} and \texttt{CRAFT$_{\text{Qwen3}}$}, respectively. These models were selected to explore trade-offs between attack performance and generation quality, offering comprehensive characteristics for adversarial generation.
 
We fine-tuned both models with maximum likelihood estimation using the diamond dataset, formatted as input–output pairs \((q, \mathcal{C}_q, d, s)\). For each triplet \((q, \mathcal{C}_q, d)\), the target adversarial perturbation $s$ was used as the supervision signal. This enabled the models to capture distributional patterns of effective adversarial insertions. Training was performed with mixed precision and gradient accumulation to support large batch sizes, and early stopping was applied to prevent overfitting.
 
\noindent\textbf{Direct Preference Optimization.}
Finally, we applied Direct Preference Optimization (DPO)~\cite{rafailov2023direct} to align the fine-tuned models with adversarial objectives. For each query–document–context triplet, we constructed two contrastive candidates \((s^+, s^-)\): $s^+$ denoting a preferred perturbation that improved ranking performance to within the top-$k$ (with $k=10$), and $s^-$ denoting a non-preferred perturbation that failed to meet the threshold. These pairs were generated from the model's output, with ranking feedback provided directly by the neural ranker (msmarco-MiniLM-L-12-v2~\cite{sbert}).

\noindent\textbf{Computational Efficiency.} All experiments were conducted on a server equipped with an 8-core CPU, 128~GB of system memory, and four NVIDIA RTX~6000 Ada GPUs, each with 48~GB of VRAM. For the supervised fine-tuning (SFT) and direct preference optimization (DPO) stages, training was distributed across four GPUs to accommodate large language models and enable efficient gradient based optimization. During inference, we employ quantized versions of the fine-tuned models and use a single GPU, as inference does not require multi-GPU parallelism. This design substantially reduces inference time, memory, and compute requirements while preserving attack effectiveness.

\vspace{-1.0em}
\section{Results and Findings}
\label{sec:results}

To evaluate the effectiveness of the proposed \texttt{CRAFT} framework, we organize our analysis around a set of research questions (RQs). These RQs are designed to collectively address the objectives $O_{[1-3]}$ as well as the threshold criteria defined in Equation~\ref{eq:rank}:

\begin{itemize}
    \item \textbf{RQ1:} How effective is the RAG-based adversarial dataset generation method in generating high-quality supervised training examples?

    \item \textbf{RQ2:} How does \texttt{CRAFT} compare to existing state-of-the-art adversarial attack baselines in terms of attack performance and content fidelity?

    \item \textbf{RQ3:} Do adversarial perturbations generated by \texttt{CRAFT} generalize effectively across different neural ranking architectures, and how does this transferability compare to that achieved by existing baselines?

    \item \textbf{RQ4:} How do the individual training stages and training data volume contribute to the attack effectiveness of \texttt{CRAFT}?

    \item \textbf{RQ5:} How well do adversarial documents generated by \texttt{CRAFT} preserve linguistic coherence, fluency, and grammaticality compared to existing baselines?

    \item \textbf{RQ6:} To what extent can adversarial documents generated by \texttt{CRAFT} evade adversarial detection filters and automated spam detection tools compared to existing baselines?

\end{itemize}

\subsection{Evaluation of Supervised Adversarial Datasets}
\label{sec:datasets_evaluation}
To investigate \textbf{RQ1}, we evaluate the performance of generated adversarial datasets, gold and diamond, produced by the approach introduced in Section~\ref{sec:dataset_generation} to determine their effectiveness as supervised training resources for adversarial attack methods. Tables~\ref{tab:gold_dataset_evaluation} and~\ref{tab:diamond_dataset_evaluation} summarize the evaluation results for the Gold and Diamond datasets, respectively. Our analysis is organized into three perspectives: (i) attack performance, measuring the extent to which adversarial documents improve the rank of target documents, (ii) linguistic acceptability score (AcS), which captures the quality and readability, and (iii) semantic similarity (SS) that captures content fidelity to evaluate the generated perturbations. To provide a clear comparison, we present and discuss the results for the Gold and Diamond datasets separately.

\begin{table}[tbp]
\centering
\small
\setlength{\tabcolsep}{4pt}
\caption{Evaluation of the Gold dataset on Easy-5 and Hard-5 groups. Attack effectiveness and linguistic quality are reported for adversarial documents across different insertion positions, with comparisons to the original documents.}
\label{tab:gold_dataset_evaluation}

\begin{tabular}{llllccccccc}
\toprule
Target  Group & Document  Type & Sentence Position 
  & Count & ASR & Top-10
  & Top-50 & Boost 
  & PPL$\downarrow$ & AcS & SS\\
\midrule
\multirow{7}{*}{Easy-5}
  & Original & –       & 5,000 & –    & –    & –    & –     & 37.3 & 0.78 & - \\ \cmidrule{2-11}
  & Adversarial  & All     & 5,000 & 99.7 & 62.2 & 93.1 & 59.2  & 42.4 & 0.77 & 0.88 \\ \cmidrule{3-11}
  &       & v=0     & 3,134 &100.0 & 72.9 & 96.6 & 63.2  & 42.2 & 0.77 & 0.87 \\
  &       & v=1     &   961 & 99.7 & 54.4 & 92.5 & 57.7  & 42.7 & 0.77 & 0.88 \\
  &       & v=2     &   371 & 99.7 & 44.5 & 87.9 & 51.6  & 40.0 & 0.76 & 0.89 \\
  &       & v=3     &   163 & 99.4 & 33.1 & 85.9 & 50.6  & 40.4 & 0.73 & 0.89 \\
  &       & v$\geq$4&   371 & 98.1 & 22.4 & 72.5 & 40.5  & 46.5 & 0.77 & 0.88 \\
\midrule
\multirow{7}{*}{Hard-5}
  & Original & –       & 5,000 & –    & –    & –    & –     & 51.4 & 0.72 & - \\ \cmidrule{2-11}
  & Adversarial  & All     & 5,000 & 99.9 & 29.1 & 49.4 & 781.1 & 67.4 & 0.72 & 0.85 \\ \cmidrule{3-11}
  &       & v=0     & 3,281 & 99.9 & 34.5 & 55.0 & 802.5 & 66.9 & 0.73 & 0.85 \\
  &       & v=1     &   854 &100.0 & 25.3 & 44.7 & 773.3 & 67.2 & 0.69 & 0.86 \\
  &       & v=2     &   341 &100.0 & 14.1 & 36.1 & 726.9 & 66.1 & 0.70 & 0.87 \\
  &       & v=3     &   142 &100.0 &  7.7 & 28.9 & 701.6 & 64.6 & 0.63 & 0.87 \\
  &       & v$\geq$4&   382 &100.0 & 12.8 & 31.2 & 692.3 & 74.0 & 0.72 & 0.84 \\
\bottomrule
\end{tabular}
\end{table}

\noindent\textbf{Gold Dataset Performance.}
Table~\ref{tab:gold_dataset_evaluation} reports the performance of adversarial examples in the Gold dataset across both Easy-5 and Hard-5 target groups. For the Easy-5 group, adversarial documents achieve substantial rank promotion boosting 62.2\% of documents into Top-10 and 93.1\% of documents into Top-50, with an average Boost of 59.2 positions. The position of the injected adversarial sentence has a clear effect. Insertions at the beginning of the document (v=0) are most effective, achieving 72.9\% Top-10, 96.6\% Top-50, and an average Boost of 63.2. Performance gradually reduces with later insertions, having v$\geq$4 achieving only 22.4\% Top-10 and an average Boost of 40.5, though ASR remains high at 98.1\%. For the Hard-5 group, the attack task is significantly more challenging due to the lower initial rank of the target documents. Nonetheless, adversarial perturbations remain highly effective, with 29.1\% of documents boosted into Top-10, 49.4\% boosted into Top-50, and an extraordinary average Boost of 781.1 rank positions. Similar to Easy-5 target group, early insertions (v=0) produce the strongest results, delivering 34.5\% Top-10 promotion and an average Boost exceeding 800 positions, while later insertions (v$\geq$3) show reduced gains.  

In terms of linguistic quality, adversarial examples remain fluent and natural. For Easy-5, perplexity (PPL) increases modestly from 37.3 over original documents to 42.4 over adversarial documents, while the acceptability score (AcS) remains stable. A similar pattern is observed in Hard-5, where perplexity rises from 51.4 to 67.4, yet the acceptability score remains unchanged at 0.72. Importantly, semantic similarity (SS) remains consistently high across insertion strategies, confirming that adversarial edits preserve the core meaning of the payload. A similar pattern is observed in Hard-5, where perplexity increases from 51.4 to 67.4, yet both acceptability score and semantic similarity remain stable, indicating minimal semantic drift. These results indicate that the Gold dataset achieves a strong balance by achieving substantial improvements in ranking effectiveness without sacrificing linguistic quality or content fidelity, validating its suitability as a supervised training resource for adversarial attack models.

\begin{table}[tbp]
\centering
\small
\setlength{\tabcolsep}{7pt}
\caption{Evaluation of the Diamond dataset on Easy-5 and Hard-5 groups. Attack effectiveness and linguistic quality are reported for adversarial documents across different insertion positions, with comparisons to the original documents.}

\label{tab:diamond_dataset_evaluation}
\scalebox{0.98}{
\begin{tabular}{llllccccc}
\toprule
Target Group & Document Type & Sentence Position
  & Counts & Top-10
  & Boost
  & PPL$\downarrow$ & AcS & SS \\
\midrule
\multirow{7}{*}{Easy-5}
  & Original    & —       & 3,110 & –    & –      & 37.6 & 0.78 & - \\ \cmidrule{2-9}
  & Adversarial     & all     & 3,110 & 100.0 & 69.7  & 41.9 & 0.77 & 0.88 \\ \cmidrule{3-9}
  &          & v=0     & 2,285 & 100.0 & 69.7  & 41.3 & 0.78 & 0.88 \\
  &          & v=1     &   523 & 100.0 & 69.7  & 44.8 & 0.76 & 0.88 \\
  &          & v=2     &   165 & 100.0 & 70.5  & 39.3 & 0.74 & 0.89 \\
  &          & v=3     &    54 & 100.0 & 67.3  & 38.4 & 0.75 & 0.90 \\
  &          & v$\geq$4&    83 & 100.0 & 70.2  & 48.7 & 0.74 & 0.88 \\
\midrule
\multirow{7}{*}{Hard-5}
  & Original    & —       & 2,470 & –    & –      & 55.8 & 0.71 & - \\ \cmidrule{2-9}
  & Adversarial     & all     & 2,470 & 58.9 & 983.1  & 73.1 & 0.71 & 0.84 \\ \cmidrule{3-9}
  &          & v=0     & 1,805 & 62.7 & 984.0  & 70.8 & 0.72 & 0.84 \\
  &          & v=1     &   382 & 56.5 & 982.3  & 79.9 & 0.66 & 0.85 \\
  &          & v=2     &   123 & 39.0 & 979.3  & 74.8 & 0.71 & 0.86 \\
  &          & v=3     &    41 & 26.8 & 974.9  & 76.2 & 0.63 & 0.87 \\
  &          & v$\geq$4&   119 & 41.2 & 979.3  & 82.0 & 0.69 & 0.82 \\
\bottomrule
\end{tabular}}
\end{table}

\noindent\textbf{Diamond Dataset Performance.}
Table~\ref{tab:diamond_dataset_evaluation} reports the results for the Diamond dataset, which applies a stricter filtering criterion by retaining only those adversarial examples that achieve substantial rank promotion, namely Easy-5 targets promoted into the Top-10 and Hard-5 targets promoted into the Top-50. This process reduces dataset size relative to Gold but yields higher-quality and more reliable adversarial examples.  

For the Easy-5 group, the results confirm the high effectiveness of the retained adversarial perturbations. Across all insertion positions, adversarial documents achieve a 100.0\% Top-10 rate, validating the consistency of the Diamond filtering process. The average Boost is 69.7 ranks, which represents an improvement over the Gold dataset average of 59.2. Performance remains stable across different insertion positions: v=0 achieves a Boost of 69.7, v=2 yields 70.5, and v$\geq$4 achieves 70.2, indicating that the Diamond selection ensures uniformly strong rank promotion regardless of sentence placement. For the Hard-5 group, the Diamond dataset captures the most successful adversarial perturbations from an inherently more challenging setting. Overall, 58.9\% of adversarial documents are promoted into the Top-10, with an extraordinary average Boost of 983.1 ranks. Early insertions again prove most effective, with v=0 achieving 62.7\% Top-10 promotion and an average Boost of 984.0 ranks. Later insertions demonstrate a marked decline in effectiveness, with v=2 and v=3 reaching only 39.0\% and 26.8\% Top-10 promotion, respectively, despite retaining high Boost values. This illustrates the greater sensitivity of Hard-5 adversarial success to insertion position.  

With respect to perturbations quality, adversarial examples in the Diamond dataset maintain high acceptability scores and low perplexity scores. For Easy-5, perplexity increases moderately from 37.6 to 41.9, while the acceptability score remains nearly unchanged. For Hard-5, perplexity rises from 55.8 to 73.1, but acceptability scores again remain stable at 0.71. Importantly, in terms of semantic similarity Easy-5 adversarial documents achieve an score of 0.88–0.90 across insertion positions, and Hard-5 achieves 0.82–0.87. These values indicate that despite aggressive rank promotion, adversarial documents preserve strong semantic alignment with their original counterparts.

The Diamond dataset provides adversarial examples that combine high attack effectiveness with preserved linguistic quality, making it an ideal training resource. By learning from these carefully filtered instances, LLMs can internalize realistic, transferable strategies for generating adversarial perturbations that satisfy both attack and quality constraints which are precisely the objectives of our proposed \texttt{CRAFT} framework.

\vspace{-1.0em}
\subsection{Comparative Attack Performance Analysis}
\label{sec:attack_performance_comparison}
\begin{table*}[tbp]
\centering
\setlength{\tabcolsep}{10pt}
\caption{Attack performance and content fidelity of \texttt{CRAFT} compared to baseline attack methods on Easy-5 and Hard-5 target groups on the MS MARCO Dev dataset. The best attack performance in each column is highlighted and ${\dagger}$ indicates statistically significant attack performance improvements over the best performing baseline, based on a paired two-tailed t-test ($p < 0.05$).}
\scalebox{0.85}{
\begin{tabular}{lcccccccc}
\toprule \toprule

\multicolumn{9}{c}{\textbf{MS MARCO Dev}} \\
\midrule 

\multicolumn{9}{c}{\textbf{Easy-5}} \\
\midrule
\addlinespace[2pt]
\multicolumn{1}{c}{} & \multicolumn{4}{c}{Attack Performance} & \multicolumn{4}{c}{Content Fidelity} \\
\cmidrule(lr){2-5}\cmidrule(lr){6-9}
Method & ASR & Top-10 & Top-50 & Boost & SS$\uparrow$ & ATI$\downarrow$ & ADT$\downarrow$ & LOR$\uparrow$ \\
\midrule
\texttt{PRADA}        & 59.8 & 1.2  & 25.2 & 13.4 & 0.9 & 0.1  & 13.1 & 0.9 \\
\texttt{Brittle-BERT} & 76.3 & 12.9 & 56.8 & 22.6 & 0.9 & 11.6 & 11.6 & 1.0 \\
\texttt{PAT}          & 46.8 & 1.4  & 17.2 & -3.3 & 0.9 & 6.3  & 6.3  & 1.0 \\
\texttt{IDEM}         & 97.3 & 32.1 & 84.8 & 49.3 & 0.9 & 11.6 & 11.6 & 1.0 \\
\texttt{EMPRA}        & \best{99.4} & 43.5 & 93.4 & 57.6 & 0.9 & 29.8 & 29.8 & 1.0 \\
\texttt{AttChain}     & 92.1 & 34.5 & 83.9 & 47.9 & 0.8 & 22.4 & 38.8 & 0.9 \\
\midrule
\texttt{CRAFT$_{\text{Qwen3}}$}    & 97.2 & 37.0 & 91.4 & 54.5 & 0.9 & 19.1 & 19.1 & 1.0 \\
\texttt{CRAFT$_{\text{Llama3.3}}$} & \best{99.4} & \best{44.5} & \best{95.8}$^{\dagger}$ & \best{59.7}$^{\dagger}$ & 0.9 & 19.9 & 19.9 & 1.0 \\
\midrule
\addlinespace[7pt]
\toprule
\multicolumn{9}{c}{\textbf{Hard-5}} \\
\midrule
\addlinespace[2pt]
\multicolumn{1}{c}{} & \multicolumn{4}{c}{Attack Performance} & \multicolumn{4}{c}{Content Fidelity} \\
\cmidrule(lr){2-5}\cmidrule(lr){6-9}
Method & ASR & Top-10 & Top-50 & Boost & SS$\uparrow$ & ATI$\downarrow$ & ADT$\downarrow$ & LOR$\uparrow$ \\
\midrule
\texttt{PRADA}        & 74.3 & 0.0  & 0.0  & 75.5 & 0.9 & 0.1  & 18.5 & 0.9 \\
\texttt{Brittle-BERT} & 99.7 & 4.2  & 23.4 & 744.5 & 0.9 & 11.2 & 11.3 & 1.0 \\
\texttt{PAT}          & 80.1 & 0.1  & 0.4  & 79.6 & 0.9 & 11.2 & 6.3  & 1.0 \\
\texttt{IDEM}         & 99.8 & 8.3  & 34.5 & 780.8 & 0.9 & 11.2 & 22.4 & 1.0 \\
\texttt{EMPRA}        & 99.3 & 10.7 & 40.8 & 828.5 & 0.8 & 32.7 & 32.7 & 1.0 \\
\texttt{AttChain}     & 99.8 & 12.2 & 42.4 & 855.2 & 0.7 & 22.8 & 39.0 & 0.9 \\
\midrule
\texttt{CRAFT$_{\text{Qwen3}}$}    & \best{100.0} & 15.3$^{\dagger}$ & 57.1$^{\dagger}$ & 911.5$^{\dagger}$ & 0.8 & 19.1 & 19.1 & 1.0 \\
\texttt{CRAFT$_{\text{Llama3.3}}$} & \best{100.0} & \best{22.2}$^{\dagger}$ & \best{70.5}$^{\dagger}$ & \best{940.5}$^{\dagger}$ & 0.8 & 19.7 & 19.7 & 1.0 \\
\bottomrule  \bottomrule
\end{tabular}
}
\label{tab:attack_eval}
\end{table*}

\begin{table*}[tbp]
\centering
\setlength{\tabcolsep}{10pt}
\caption{Attack performance and content fidelity of \texttt{CRAFT} compared to baseline attack methods on Easy-5 and Hard-5 target groups on the TREC DL 2019 dataset. The best attack performance in each column is highlighted and ${\dagger}$ indicates statistically significant attack performance improvements over the best performing baseline, based on a paired two-tailed t-test ($p < 0.05$).}
\label{tab:attack_eval_2019}
\scalebox{0.85}{
\begin{tabular}{lcccccccc}
\toprule \toprule
\multicolumn{9}{c}{\textbf{TREC DL 2019}} \\
\midrule 
\multicolumn{9}{c}{\textbf{Easy-5}} \\
\midrule
\addlinespace[2pt]
\multicolumn{1}{c}{} & \multicolumn{4}{c}{Attack Performance} & \multicolumn{4}{c}{Content Fidelity} \\
\cmidrule(lr){2-5}\cmidrule(lr){6-9}
Method & ASR & Top-10 & Top-50 & Boost & SS$\uparrow$ & ATI$\downarrow$ & ADT$\downarrow$ & LOR$\uparrow$ \\
\midrule
\texttt{PRADA}        & 73.1 & 1.6  & 39.4 & 19.7 & 0.9 & 0.1  & 14.5 & 0.9 \\
\texttt{Brittle-BERT} & 87.5 & 42.1 & 78.9 & 44.9 & 0.9 & 11.5 & 11.5 & 1.0 \\
\texttt{PAT}          & 51.1 & 3.6  & 22.7 & -0.4 & 0.9 & 6.4  & 6.4  & 1.0 \\
\texttt{IDEM}         & 98.1 & 63.9 & 92.9 & 60.6 & 0.9 & 11.5 & 11.5 & 1.0 \\
\texttt{EMPRA}        & 99.8 & 78.7 & 97.3 & 67.1 & 0.8 & 27.4 & 27.4 & 1.0 \\
\texttt{AttChain}     & 97.8 & 64.8 & 94.2 & 62.0 & 0.8 & 18.3 & 31.9 & 0.9 \\
\midrule	
\texttt{CRAFT$_{\text{Qwen3}}$}    & 99.1 & 76.1 & 97.3 & 65.6 & 0.9 & 23.6 & 23.6 & 1.0 \\
\texttt{CRAFT$_{\text{Llama3.3}}$} & \best{99.9} & \best{79.5} & \best{98.4} & \best{67.6} & 0.9 & 19.8 & 19.8 & 1.0 \\
\midrule
\addlinespace[7pt]
\toprule
\multicolumn{9}{c}{\textbf{Hard-5}} \\
\midrule
\addlinespace[2pt]
\multicolumn{1}{c}{} & \multicolumn{4}{c}{Attack Performance} & \multicolumn{4}{c}{Content Fidelity} \\
\cmidrule(lr){2-5}\cmidrule(lr){6-9}
Method & ASR & Top-10 & Top-50 & Boost & SS$\uparrow$ & ATI$\downarrow$ & ADT$\downarrow$ & LOR$\uparrow$ \\
\midrule
\texttt{PRADA}        & 75.6 & 0.0  & 0.2  & 85.5  & 0.8 & 0.1  & 16.5 & 0.8 \\
\texttt{Brittle-BERT} & 99.9 & 16.1 & 47.6 & 848.3 & 0.8 & 11.2 & 11.2 & 1.0 \\
\texttt{PAT}          & 80.5 & 0.2  & 0.5  & 110.1 & 0.9 & 6.5 & 6.5  & 1.0 \\
\texttt{IDEM}         & 99.8 & 26.2 & 54.3 & 842.3 & 0.9 & 11.2 & 11.2 & 1.0 \\
\texttt{EMPRA}        & \best{100.0} & 44.2 & 74.4 & 943.2 & 0.8 & 32.1 & 32.2 & 1.0 \\
\texttt{AttChain}     & 99.9 & 35.0 & 67.6 & 916.9 & 0.7 & 19.7 & 32.0 & 0.9 \\
\midrule
\texttt{CRAFT$_{\text{Qwen3}}$}    & \best{100.0} & 52.0$^{\dagger}$ & 80.7$^{\dagger}$ & 951.3$^{\dagger}$ & 0.8 & 24.3 & 24.3 & 1.0 \\
\texttt{CRAFT$_{\text{Llama3.3}}$} & \best{100.0} & \best{56.9}$^{\dagger}$ & \best{84.1}$^{\dagger}$ & \best{958.2}$^{\dagger}$ & 0.8 & 20.4 & 20.4 & 1.0 \\
\bottomrule  \bottomrule
\end{tabular}
}
\end{table*}

\begin{table*}[tbp]
\centering
\setlength{\tabcolsep}{10pt}
\caption{Attack performance and content fidelity of \texttt{CRAFT} compared to baseline attack methods on Easy-5 and Hard-5 target groups on the TREC DL 2020 dataset. The best attack performance in each column is highlighted and ${\dagger}$ indicates statistically significant attack performance improvements over the best performing baseline, based on a paired two-tailed t-test ($p < 0.05$).}
\label{tab:attack_eval_2020}
\scalebox{0.85}{
\begin{tabular}{lcccccccc}
\toprule \toprule
\multicolumn{9}{c}{\textbf{TREC DL 2020}} \\
\midrule 
\multicolumn{9}{c}{\textbf{Easy-5}} \\
\midrule
\addlinespace[2pt]
\multicolumn{1}{c}{} & \multicolumn{4}{c}{Attack Performance} & \multicolumn{4}{c}{Content Fidelity} \\
\cmidrule(lr){2-5}\cmidrule(lr){6-9}
Method & ASR & Top-10 & Top-50 & Boost & SS$\uparrow$ & ATI$\downarrow$ & ADT$\downarrow$ & LOR$\uparrow$ \\
\midrule
\texttt{PRADA}        & 73.2 & 1.7  & 37.7 & 19.3 & 0.9 & 0.0 & 14.9 & 0.9 \\
\texttt{Brittle-BERT} & 86.5 & 42.2 & 77.6 & 44.0 & 0.9 & 11.6 & 11.6 & 1.0 \\
\texttt{PAT}          & 53.2 & 2.3  & 23.6 & 2.7  & 0.9 & 6.3 & 6.3 & 1.0 \\
\texttt{IDEM}         & 98.4 & 60.4 & 91.3 & 59.6 & 0.9 & 11.5 & 11.5 & 1.0 \\
\texttt{EMPRA}        & 99.8 & 78.3 & \best{98.5} & \best{67.5} & 0.9 & 26.8 & 26.8 & 1.0 \\
\texttt{AttChain}     & 97.7 & 63.1 & 93.8 & 60.6 & 0.8 & 18.6 & 31.7 & 0.9 \\
\midrule
\texttt{CRAFT$_{\text{Qwen3}}$}    & 98.4 & 75.8 & 95.5 & 65.1 & 0.9 & 22.3 & 22.3 & 1.0 \\
\texttt{CRAFT$_{\text{Llama3.3}}$} & \best{99.6} & \best{79.7} & 98.0 & \best{67.5} & 0.9 & 19.9 & 19.9 & 1.0 \\
\midrule
\addlinespace[7pt]
\toprule
\multicolumn{9}{c}{\textbf{Hard-5}} \\
\midrule
\addlinespace[2pt]
\multicolumn{1}{c}{} & \multicolumn{4}{c}{Attack Performance} & \multicolumn{4}{c}{Content Fidelity} \\
\cmidrule(lr){2-5}\cmidrule(lr){6-9}
Method & ASR & Top-10 & Top-50 & Boost & SS$\uparrow$ & ATI$\downarrow$ & ADT$\downarrow$ & LOR$\uparrow$ \\
\midrule
\texttt{PRADA}        & 74.9  & 0.0  & 0.1  & 80.4  & 0.9 & 0.1 & 16.3 & 0.8 \\
\texttt{Brittle-BERT} & \best{100.0} & 20.1 & 47.8 & 834.8 & 0.8 & 11.3 & 11.3 & 1.0 \\
\texttt{PAT}          & 80.8  & 0.0  & 0.4  & 112.2 & 0.9 & 6.4 & 6.4 & 1.0 \\
\texttt{IDEM}         & 99.9  & 25.5 & 53.1 & 854.9 & 0.9 & 11.3 & 11.4 & 1.0 \\
\texttt{EMPRA}        & 99.9  & 44.6 & 71.6 & 930.2 & 0.8 & 29.0 & 29.1 & 1.0 \\
\texttt{AttChain}     & 99.7 & 35.4 & 66.7 & 915.6 & 0.8 & 21.3 & 32.7 & 0.9 \\
\midrule
\texttt{CRAFT$_{\text{Qwen3}}$}    & 99.8 & 50.2$^{\dagger}$ & 78.2$^{\dagger}$ & 948.2$^{\dagger}$ & 0.8 & 23.6 & 23.6 & 1.0 \\
\texttt{CRAFT$_{\text{Llama3.3}}$} & \best{100.0} & \best{53.8}$^{\dagger}$ & \best{81.4}$^{\dagger}$ & \best{960.3}$^{\dagger}$ & 0.8 & 19.9 & 19.9 & 1.0 \\
\bottomrule  \bottomrule
\end{tabular}
}

\end{table*}

To explore \textbf{RQ2}, we evaluate the attack performance of the proposed \texttt{CRAFT} framework, trained on our supervised dataset, in comparison with state-of-the-art adversarial attack baselines across both the Easy-5 and Hard-5 target groups. Tables~\ref{tab:attack_eval}--\ref{tab:attack_eval_2020} report results across four attack performance metrics along with content fidelity measures on MS MARCO Dev, TREC DL 2019, and TREC DL 2020. These metrics jointly capture the ability of adversarial methods to promote documents in the ranking while preserving semantic and stylistic fidelity of the original document.

On MS MARCO Dev, for the Easy-5 group, \texttt{CRAFT$_{\text{Llama3.3}}$} achieves the strongest overall performance, significantly outperforming all baselines with a Top-10 success rate of 44.5\%, a Top-50 success rate of 95.8\%, and the highest Boost of 59.7 ranks. Compared to the strongest baseline for this group, \texttt{EMPRA}, which records 43.5\% Top-10 and 93.4\% Top-50, \texttt{CRAFT} delivers consistent improvements while preserving fidelity, with semantic similarity at 0.9 and ATI at 19.0. Importantly, these gains in Top-50 and Boost are statistically significant relative to the best baseline, highlighting the robustness of \texttt{CRAFT}'s improvements. Among other baselines, \texttt{IDEM} achieves relatively strong promotion with 32.1\% Top-10 and 84.8\% Top-50, though it lags behind both \texttt{EMPRA} and \texttt{CRAFT}. \texttt{AttChain} shows competitive Top-10 and Top-50 scores but suffers from higher content distortion (ADT 38.8). Earlier approaches such as \texttt{PRADA}, \texttt{PAT}, and \texttt{Brittle-BERT} perform substantially worse, with very low Top-10 rates, all below 13\%, confirming their limited effectiveness in the Easy-5 setting.

These findings are consistent across the TREC DL benchmarks. As shown in Tables~\ref{tab:attack_eval_2019} and~\ref{tab:attack_eval_2020}, \texttt{CRAFT$_{\text{Llama3.3}}$} achieves the highest or near-highest Easy-5 attack performance on both TREC DL 2019 and TREC DL 2020, with Top-10 rates of 79.5\% and 79.7\%, respectively. \texttt{EMPRA} remains competitive in this setting, achieving comparable Top-50 and Boost scores, and even a marginally higher Top-50 of 98.5\% on TREC DL 2020. This competitive attack performance comes at the cost of lower content fidelity, with \texttt{EMPRA} recording SS of 0.8 on both TREC DL benchmarks compared to 0.9 for \texttt{CRAFT$_{\text{Llama3.3}}$}, alongside substantially higher ATI values of 27.4 and 26.8 versus 19.8 and 19.9. The consistency of these results across three benchmarks with diverse query topics confirms that the adversarial generation capabilities of \texttt{CRAFT} generalize effectively across different query distributions and evaluation settings.

On MS MARCO Dev, for the more challenging Hard-5 group, \texttt{CRAFT} demonstrates an even greater margin of improvement. Both \texttt{CRAFT$_{\text{Qwen3}}$} and \texttt{CRAFT$_{\text{Llama3.3}}$} achieve perfect ASR of 100.0\%. \texttt{CRAFT$_{\text{Llama3.3}}$} reaches the highest Top-10 rate of 22.2\% and Top-50 rate of 70.5\%, significantly surpassing the best baseline \texttt{AttChain}, which records 12.2\% Top-10 and 42.4\% Top-50. The Boost achieved by \texttt{CRAFT$_{\text{Llama3.3}}$} is 940.5 ranks, clearly exceeding the baseline maximum of 855.2. While \texttt{EMPRA} and \texttt{IDEM} achieve moderate promotion with 10.7 and 8.3\% Top-10 and 40.8 and 34.5\% Top-50 respectively, their performance falls far short of \texttt{CRAFT}. Earlier baselines such as \texttt{Brittle-BERT}, \texttt{PAT}, and \texttt{PRADA} struggle considerably in this setting, with Top-10 scores below 5\%. Importantly, across all methods, \texttt{CRAFT} maintains strong semantic fidelity with SS of 0.8 and ATI of 19.1, avoiding the quality degradation observed in \texttt{EMPRA}, which records ATI of 32.7 and ADT of 32.7, or \texttt{AttChain}, which records ATI of 22.8 and ADT of 39.0. It is important to note that baselines such as \texttt{CRAFT} and \texttt{IDEM} exhibit comparable ATI and ADT values because they do not alter the core content of the document. By contrast, insertion-heavy approaches such as \texttt{PRADA} and \texttt{AttChain} introduce extensive edits to the document content, resulting in substantially higher distortion scores and causing drift from the core meaning of the original text.

The advantage of \texttt{CRAFT} in the Hard-5 setting is further amplified on the TREC DL benchmarks. On both TREC DL 2019 and 2020, \texttt{CRAFT$_{\text{Llama3.3}}$} achieves Top-10 rates of 56.9\% and 53.8\% and Top-50 rates of 84.1\% and 81.4\%, respectively, significantly outperforming all baselines across every metric. Relative to \texttt{EMPRA}, the strongest baseline in this setting on TREC DL, \texttt{CRAFT$_{\text{Llama3.3}}$} achieves improvements of 28.7\% and 20.6\% in Top-10 and 13.0\% and 13.7\% in Top-50 on TREC DL 2019 and 2020, respectively, with all gains being statistically significant. Notably, while \texttt{AttChain} leads the Hard-5 baselines on MS MARCO Dev, \texttt{EMPRA} takes the lead on both TREC DL benchmarks, indicating that the relative ranking of baseline methods shifts across query distributions. In contrast, \texttt{CRAFT} consistently achieves the highest performance regardless of the dataset, suggesting that its preference-optimized training pipeline produces adversarial content that generalizes effectively to unseen query sets with diverse information needs. Content fidelity remains stable across all three benchmarks, with \texttt{CRAFT} consistently preserving SS of 0.8 while \texttt{EMPRA} and \texttt{AttChain} drop to SS of 0.7--0.8 with notably higher distortion scores.

\vspace{-1.0em}

\subsection{Cross-Model Transferability of Adversarial Perturbations}
\label{sec:cross_nrm_attack_performance_comparison}

\begin{table*}[t]
\centering
\setlength{\tabcolsep}{7pt}
\caption{Attack performance of adversarial documents generated by \texttt{CRAFT} and baselines on Easy-5 and Hard-5 across different victim NRMs on the MS MARCO Dev dataset. The best attack performance in each column is highlighted, and ${\dagger}$ indicates statistically significant improvements over the best performing baseline according to a paired two-tailed t-test ($p<0.05$).}
\label{tbl:victim_models_results}

\scalebox{0.8}{
\begin{tabular}{lcccccccc}
\toprule \toprule

\multicolumn{9}{c}{\textbf{MS MARCO Dev}} \\
\midrule 

 Method & \multicolumn{4}{c}{Easy-5} & \multicolumn{4}{c}{Hard-5} \\ 
\cmidrule{1-5} \cmidrule(lr){6-9}
\textbf{DistilRoBERta}  & ASR & Top-10 & Top-50 & Boost & ASR & Top-10 & Top-50 & Boost\\
\midrule

\texttt{PRADA}        & 61.1 & 2.7  & 24.4 & 23.1 & 59.4 & 0.0  & 0.1  & 30.8 \\
\texttt{Brittle-BERT} & 76.3 & 16.1 & 54.4 & 65.9 & 95.4 & 6.2  & 22.1 & 524.6 \\
\texttt{PAT}          & 49.3 & 1.0  & 18.3 & 4.9  & 61.6 & 0.1  & 0.5  & 52.7 \\
\texttt{IDEM}         & 93.0 & 31.2 & 73.6 & 94.7 & 96.1 & 11.3 & 33.9 & 565.0 \\
\texttt{EMPRA}        & 95.9 & 37.1 & 83.6 & 104.1& 94.6 & 15.4 & 40.1 & 596.7 \\
\texttt{AttChain}     & 93.0 & 35.1 & 77.7 & 101.2& 98.7 & 14.4 & 40.5 & 637.6 \\
\midrule
\texttt{CRAFT$_{\text{Qwen3}}$}    & 95.6 & 39.2 & 83.1 & \best{108.1} & \best{99.6}$^{\dagger}$ & 20.8$^{\dagger}$ & 57.7$^{\dagger}$ & 698.9$^{\dagger}$ \\
\texttt{CRAFT$_{\text{Llama3.3}}$} & \best{97.6} & \best{43.1}$^{\dagger}$ & \best{89.6}$^{\dagger}$ & \best{113.5}$^{\dagger}$ & 99.5 & \best{23.5}$^{\dagger}$ & \best{59.9}$^{\dagger}$ & \best{711.6}$^{\dagger}$ \\

\midrule

\textbf{ELECTRA} & ASR & Top-10 & Top-50 & Boost & ASR & Top-10 & Top-50 & Boost\\
\midrule
\texttt{PRADA}        & 42.3 & 2.0  & 24.1 & 15.4 & 34.4 & 0.0  & 0.0  & -1.3 \\
\texttt{Brittle-BERT} & 73.8 & 18.1 & 55.9 & 55.1 & 95.9 & 6.3  & 25.9 & 490.5 \\
\texttt{PAT}          & 50.1 & 1.1  & 20.9 & 7.0  & 40.6 & 0.0  & 0.4  & -22.4 \\
\texttt{IDEM}         & 93.8 & 30.7 & 77.4 & 85.6 & 94.5 & 9.6  & 32.2 & 497.2 \\
\texttt{EMPRA}        & 97.0 & 42.9 & 88.2 & 95.6 & 94.9 & 14.6 & 39.3 & 557.7 \\
\texttt{AttChain}     & 91.2 & 34.7 & 78.3 & 85.2 & 97.7 & 11.7 & 40.3 & 579.8 \\
\midrule
\texttt{CRAFT$_{\text{Qwen3}}$} & 95.5 & 36.8 & 86.2 & 94.8 & \best{99.8}$^{\dagger}$ & \best{17.5}$^{\dagger}$ & 52.5$^{\dagger}$ & 644.2$^{\dagger}$ \\
\texttt{CRAFT$_{\text{Llama3.3}}$} & \best{97.7} & \best{43.1} & \best{92.1}$^{\dagger}$ & \best{100.6} & \best{99.8}$^{\dagger}$ & 17.0$^{\dagger}$ & \best{57.8}$^{\dagger}$& \best{658.2}$^{\dagger}$ \\

\midrule

\textbf{Qwen3-Embedding} & ASR & Top-10 & Top-50 & Boost & ASR & Top-10 & Top-50 & Boost\\
\midrule
\texttt{PRADA}        & 55.7 & 1.8  & 24.6 & 14.1 & 59.1 & 0.0  & 0.0  & 35.9 \\
\texttt{Brittle-BERT} & 74.0 & 6.2  & 44.0 & 66.3 & 97.6 & 0.4  & 5.5  & 362.5 \\
\texttt{PAT}          & 57.4 & 1.1  & 21.4 & 15.0 & 79.1 & 0.0  & 0.2  & 58.5 \\
\texttt{IDEM}         & 95.7 & 16.3 & 63.8 & 103.1& 98.6 & 1.8  & 12.7 & 430.3 \\
\texttt{EMPRA}        & 97.2 & 21.7 & 72.5 & 114.0& 96.9 & 2.6  & 13.8 & 490.9 \\
\texttt{AttChain}     & 92.1 & 23.2 & 71.4 & 111.8& 99.1 & \best{6.8}  & 27.4 & 592.6 \\
\midrule
\texttt{CRAFT$_{\text{Qwen3}}$}    & 97.4 & 24.1  & 81.3$^{\dagger}$  & 126.1$^{\dagger}$  & \best{100.0}$^{\dagger}$   & 4.4  & 30.5  & 641.9$^{\dagger}$   \\
\texttt{CRAFT$_{\text{Llama3.3}}$} & \best{98.5}$^{\dagger}$  & \best{27.6}$^{\dagger}$  & \best{84.6}$^{\dagger}$  & \best{129.1}$^{\dagger}$  & \best{100.0}$^{\dagger}$   & 4.6 & \best{31.4}$^{\dagger}$   & \best{658.7}$^{\dagger}$   \\
\midrule
\textbf{RankLLM} & ASR & Top-10 & Top-50 & Boost & ASR & Top-10 & Top-50 & Boost\\
\midrule
\texttt{PRADA}        & 25.4 & 0.3  & 6.5  & -1.6 & 22.7 & 0.0  & 0.0  & -2.1 \\
\texttt{Brittle-BERT} & 27.1 & 0.6  & 9.7  & -0.5 & 79.2 & 0.1  & 2.3  & 224.1 \\
\texttt{PAT}          & 22.5 & 1.1  & 7.3  & -1.1 & 29.3 & 0.0  & 0.1  & 3.6 \\
\texttt{IDEM}         & 67.4 & 9.6  & 36.5 & 19.1 & 94.2 & 1.4  & 12.1 & 491.1 \\
\texttt{EMPRA}        & 74.3 & 13.5 & 46.1 & 25.7 & 93.7 & 2.3  & 11.1 & 511.3 \\
\texttt{AttChain}     & 75.9 & 15.9 & 48.5 & 28.4 & 97.6 & 3.6  & 20.4 & 639.0 \\
\midrule
\texttt{CRAFT$_{\text{Qwen3}}$}    & 86.0$^{\dagger}$ & 17.2 & 56.5$^{\dagger}$ & 33.8$^{\dagger}$ & 99.4$^{\dagger}$ & 4.5  & \best{27.4}$^{\dagger}$ & 725.1$^{\dagger}$ \\
\texttt{CRAFT$_{\text{Llama3.3}}$} & \best{88.7}$^{\dagger}$ & \best{19.0}$^{\dagger}$ & \best{62.2}$^{\dagger}$ & \best{37.9}$^{\dagger}$ & \best{99.6}$^{\dagger}$ & \best{5.6}$^{\dagger}$ & 25.5$^{\dagger}$ & \best{742.7}$^{\dagger}$ \\

\bottomrule
\bottomrule
\end{tabular}
}
\end{table*}

\begin{table*}[t]
    \centering
    \setlength{\tabcolsep}{7pt}
    \caption{Attack performance of adversarial documents generated by \texttt{CRAFT} and baselines on Easy-5 and Hard-5 across different victim NRMs on the TREC DL 2019 dataset. The best attack performance in each column is highlighted, and ${\dagger}$ indicates statistically significant improvements over the best performing baseline according to a paired two-tailed t-test ($p<0.05$).}
    \label{tbl:victim_models_results_2019}
    \scalebox{0.8}{

    \begin{tabular}{lcccccccc}
    \toprule \toprule
    \multicolumn{9}{c}{\textbf{TREC DL 2019}} \\
    \midrule 
     Method & \multicolumn{4}{c}{Easy-5} & \multicolumn{4}{c}{Hard-5} \\ 
    \cmidrule{1-5} \cmidrule(lr){6-9}
    \textbf{DistilRoBERta}  & ASR & Top-10 & Top-50 & Boost & ASR & Top-10 & Top-50 & Boost\\
    \midrule
    
    \texttt{PRADA}        & 56.8 & 1.9 & 23.5 & 21.8 & 57.2 & 0.0 & 0.0 & 27.9 \\
    \texttt{Brittle-BERT} & 88.3 & 36.9 & 72.5 & 111.7 & 97.5 & 16.5 & 42.9 & 600.7 \\
    \texttt{PAT}          & 51.3 & 2.9 & 21.7 & 17.6 & 63.4 & 0.0 & 0.8 & 73.5 \\
    \texttt{IDEM}         & 96.3 & 55.8 & 84.9 & 130.2 & 98.0 & 25.5 & 49.4 & 613.0 \\
    \texttt{EMPRA}        & \best{98.9} & 68.5 & 92.0 & 140.7 & 99.8 & 43.5 & 68.9 & 705.0 \\
    \texttt{AttChain}     & 96.6 & 57.1 & 89.4 & 133.5 & 99.4 & 30.5 & 59.0 & 670.4 \\
    \midrule
    \texttt{CRAFT$_{\text{Qwen3}}$}    & 98.5 & 68.8 & 92.9 & 142.4$^{\dagger}$ & 99.9 & 50.1$^{\dagger}$ & 75.6$^{\dagger}$ & 722.2$^{\dagger}$ \\
    \texttt{CRAFT$_{\text{Llama3.3}}$} & 98.7 & \best{72.3$^{\dagger}$} & \best{95.0$^{\dagger}$} & \best{144.5$^{\dagger}$} & \best{100.0} & \best{52.6$^{\dagger}$} & \best{78.0$^{\dagger}$} & \best{730.7$^{\dagger}$} \\
    
    \midrule
    
    \textbf{ELECTRA}
    & ASR & Top-10 & Top-50 & Boost
    & ASR & Top-10 & Top-50 & Boost \\
    \midrule
    \texttt{PRADA}        & 53.5 & 2.2 & 28.3 & 23.3 & 39.5 & 0.0 & 0.2 & 8.6 \\
    \texttt{Brittle-BERT} & 88.6 & 48.3 & 78.9 & 102.7 & 97.2 & 19.7 & 46.7 & 571.3 \\
    \texttt{PAT}          & 54.6 & 3.7 & 26.0 & 15.8 & 39.1 & 0.1 & 0.5 & -13.9 \\
    \texttt{IDEM}         & 96.9 & 59.6 & 89.5 & 116.4 & 93.8 & 23.0 & 47.6 & 532.8 \\
    \texttt{EMPRA}        & 99.2 & 75.1 & 96.3 & 126.8 & 99.4 & 45.6 & 69.8 & 656.5 \\
    \texttt{AttChain}     & 97.0 & 59.0 & 89.0 & 117.3 & 99.0 & 31.0 & 61.0 & 626.3 \\
    \midrule
    \texttt{CRAFT$_{\text{Qwen3}}$}    & 98.0 & 71.3 & 95.7 & 125.0 & 99.8 & 50.8$^{\dagger}$ & 76.2$^{\dagger}$ & 669.0$^{\dagger}$ \\
    \texttt{CRAFT$_{\text{Llama3.3}}$} & \best{99.2} & \best{76.2} & \best{98.0}$^{\dagger}$ & \best{127.5} & \best{100.0}$^{\dagger}$ & \best{53.7}$^{\dagger}$ & \best{78.8}$^{\dagger}$ & \best{681.1}$^{\dagger}$ \\
    \midrule
    
    \textbf{Qwen3-Embedding}
    & ASR & Top-10 & Top-50 & Boost
    & ASR & Top-10 & Top-50 & Boost \\
    \midrule
    \texttt{PRADA}        & 56.9 & 3.0 & 23.2 & 16.1 & 58.1 & 0.1 & 0.6 & 35.1 \\
    \texttt{Brittle-BERT} & 83.1 & 16.8 & 53.4 & 105.9 & 98.4 & 3.2 & 13.6 & 431.6 \\
    \texttt{PAT}          & 63.3 & 3.1 & 24.4 & 30.0 & 85.0 & 0.1 & 1.3 & 76.4 \\
    \texttt{IDEM}         & 97.8 & 31.9 & 72.3 & 140.6 & 99.4 & 8.5 & 21.7 & 469.2 \\
    \texttt{EMPRA}        & 99.2 & 49.9 & 85.3 & 164.4 & \best{100.0} & 17.5 & 39.9 & 619.6 \\
    \texttt{AttChain}     & 96.6 & 41.1 & 78.8 & 150.7 & 99.8 & 17.4 & 36.7 & 620.8 \\
    \midrule
    \texttt{CRAFT$_{\text{Qwen3}}$}    & 97.8 & 52.0 & 88.5$^{\dagger}$ & 168.9$^{\dagger}$ & 99.9 & 23.1$^{\dagger}$ & 53.4$^{\dagger}$ & 704.4$^{\dagger}$ \\
    \texttt{CRAFT$_{\text{Llama3.3}}$} & \best{99.3$^{\dagger}$} & \best{53.7$^{\dagger}$} & \best{90.1$^{\dagger}$} & \best{171.9$^{\dagger}$} & \best{100.0} & \best{26.5$^{\dagger}$} & \best{56.6$^{\dagger}$} & \best{719.0$^{\dagger}$} \\
    \midrule
    
    \textbf{RankLLM}
    & ASR & Top-10 & Top-50 & Boost 
    & ASR & Top-10 & Top-50 & Boost \\
    \midrule
    \texttt{PRADA}        & 23.9 & 0.7 & 6.2 & -1.7 & 29.4 & 0.0 & 0.0 & -1.8 \\
    \texttt{Brittle-BERT} & 40.5 & 3.7 & 17.3 & 6.4 & 87.1 & 1.5 & 8.6 & 306.8 \\
    \texttt{PAT}          & 29.3 & 0.8 & 6.3 & -0.6 & 40.2 & 0.0 & 0.1 & 7.3 \\
    \texttt{IDEM}         & 76.0 & 23.2 & 50.1 & 30.8 & 97.0 & 8.4 & 24.3 & 583.9 \\
    \texttt{EMPRA}        & 82.3 & 32.1 & 60.5 & 38.7 & 98.7 & 15.5 & 38.4 & 715.6 \\
    \texttt{AttChain}     & 83.2 & 30.9 & 60.1 & 39.5 & 99.2 & 12.6 & 33.6 & 710.3 \\
    \midrule
    \texttt{CRAFT$_{\text{Qwen3}}$}    & 88.3$^{\dagger}$ & 39.2$^{\dagger}$ & 67.9$^{\dagger}$ & 45.2$^{\dagger}$ & 99.7 & 23.4$^{\dagger}$ & \best{52.5}$^{\dagger}$ & 819.2$^{\dagger}$ \\
    \texttt{CRAFT$_{\text{Llama3.3}}$} & \best{91.0}$^{\dagger}$ & \best{41.1}$^{\dagger}$ & \best{69.6}$^{\dagger}$ & \best{47.1}$^{\dagger}$ & \best{99.8} & \best{23.5}$^{\dagger}$ & 51.6$^{\dagger}$ & \best{835.5}$^{\dagger}$ \\
    \bottomrule
    \bottomrule
    \end{tabular}
    }

    \end{table*}

\begin{table*}[t]
    \centering
    \setlength{\tabcolsep}{7pt}
    \caption{Attack performance of adversarial documents generated by \texttt{CRAFT} and baselines on Easy-5 and Hard-5 across different victim NRMs on the TREC DL 2020 dataset. The best attack performance in each column is highlighted, and ${\dagger}$ indicates statistically significant improvements over the best performing baseline according to a paired two-tailed t-test ($p<0.05$).}
    \label{tbl:victim_models_results_2020}
    \scalebox{0.8}{
    \begin{tabular}{lcccccccc}
    \toprule \toprule
    \multicolumn{9}{c}{\textbf{TREC DL 2020}} \\
    \midrule 
     Method & \multicolumn{4}{c}{Easy-5} & \multicolumn{4}{c}{Hard-5} \\ 
    \cmidrule{1-5} \cmidrule(lr){6-9}
    \textbf{DistilRoBERta}  & ASR & Top-10 & Top-50 & Boost & ASR & Top-10 & Top-50 & Boost\\
    \midrule
    
    \texttt{PRADA}        & 61.2 & 3.4 & 25.9 & 31.3 & 62.1 & 0.0 & 0.3 & 47.9 \\
    \texttt{Brittle-BERT} & 86.7 & 38.1 & 72.9 & 119.2 & 96.4 & 17.8 & 43.7 & 594.1 \\
    \texttt{PAT}          & 55.3 & 2.8 & 21.0 & 18.8 & 63.6 & 0.1 & 1.2 & 87.9 \\
    \texttt{IDEM}         & 96.6 & 55.1 & 84.3 & 134.7 & 97.7 & 24.1 & 50.2 & 620.4 \\
    \texttt{EMPRA}        & \best{98.8} & 70.0 & 93.0 & 148.9 & 99.7 & 43.3 & 65.2 & 687.6 \\
    \texttt{AttChain}     & 97.3 & 56.5 & 88.1 & 141.9 & 98.7 & 28.5 & 57.8 & 675.8 \\
    \midrule
    \texttt{CRAFT$_{\text{Qwen3}}$}    & 97.8 & 68.9 & 92.3 & 149.5 & 99.7 & 46.7 & 72.2$^{\dagger}$ & 718.1$^{\dagger}$ \\
    \texttt{CRAFT$_{\text{Llama3.3}}$} & \best{98.8} & \best{72.7} & \best{95.9$^{\dagger}$} & \best{153.1$^{\dagger}$} & \best{99.9} & \best{49.6$^{\dagger}$} & \best{76.4$^{\dagger}$} & \best{733.7$^{\dagger}$} \\
    
    \midrule
    
    \textbf{ELECTRA}
    & ASR & Top-10 & Top-50 & Boost
    & ASR & Top-10 & Top-50 & Boost \\
    \midrule
    \texttt{PRADA}        & 55.0 & 2.0 & 26.4 & 23.7 & 37.6 & 0.0 & 0.0 & 11.3 \\
    \texttt{Brittle-BERT} & 87.1 & 46.2 & 77.0 & 107.5 & 97.0 & 21.0 & 49.3 & 562.9 \\
    \texttt{PAT}          & 58.3 & 3.1 & 25.5 & 20.0 & 40.4 & 0.0 & 0.6 & -9.8 \\
    \texttt{IDEM}         & 97.3 & 55.3 & 86.3 & 119.8 & 95.8 & 21.9 & 48.7 & 554.0 \\
    \texttt{EMPRA}        & \best{99.2} & \best{73.3} & 95.1 & 131.9 & 99.4 & 44.6 & 67.4 & 643.9 \\
    \texttt{AttChain}     & 96.4 & 55.8 & 88.2 & 121.8 & 98.9 & 31.0 & 60.4 & 624.6 \\
    \midrule
    \texttt{CRAFT$_{\text{Qwen3}}$}    & 97.9 & 70.3 & 94.0 & 130.0 & \best{99.9}$^{\dagger}$ & 48.0 & 73.4$^{\dagger}$ & 661.2$^{\dagger}$ \\
    \texttt{CRAFT$_{\text{Llama3.3}}$} & 99.0 & 72.8 & \best{96.1} & \best{132.1} & \best{99.9}$^{\dagger}$ & \best{49.2}$^{\dagger}$ & \best{77.2}$^{\dagger}$ & \best{673.0}$^{\dagger}$ \\
    \midrule
    
    \textbf{Qwen3-Embedding}
    & ASR & Top-10 & Top-50 & Boost
    & ASR & Top-10 & Top-50 & Boost \\
    \midrule
    \texttt{PRADA}        & 56.2 & 3.4 & 24.4 & 11.7 & 65.1 & 0.0 & 0.5 & 52.9 \\
    \texttt{Brittle-BERT} & 83.2 & 17.4 & 58.4 & 97.7 & 99.6 & 2.3 & 14.2 & 436.4 \\
    \texttt{PAT}          & 63.8 & 2.2 & 24.8 & 24.2 & 80.3 & 0.0 & 0.5 & 82.2 \\
    \texttt{IDEM}         & 96.6 & 33.6 & 73.4 & 124.7 & 99.3 & 7.3 & 22.1 & 490.1 \\
    \texttt{EMPRA}        & \best{99.7} & 50.6 & 88.0 & 148.6 & \best{100.0} & 16.9 & 38.5 & 614.3 \\
    \texttt{AttChain}     & 95.3 & 43.3 & 77.9 & 132.8 & 99.5 & 18.3 & 41.0 & 626.0 \\
    \midrule
    \texttt{CRAFT$_{\text{Qwen3}}$}    & 97.8 & 51.4 & 90.0 & 154.3$^{\dagger}$ & \best{100.0} & 23.6$^{\dagger}$ & 52.9$^{\dagger}$ & 703.5$^{\dagger}$ \\
    \texttt{CRAFT$_{\text{Llama3.3}}$} & 99.0 & \best{56.0$^{\dagger}$} & \best{92.9$^{\dagger}$} & \best{157.1$^{\dagger}$} & \best{100.0} & \best{24.9$^{\dagger}$} & \best{53.7$^{\dagger}$} & \best{713.9$^{\dagger}$} \\
    \midrule
    
    \textbf{RankLLM}
    & ASR & Top-10 & Top-50 & Boost
    & ASR & Top-10 & Top-50 & Boost \\
    \midrule
    \texttt{PRADA}        & 24.4 & 0.9 & 6.4 & -1.9 & 25.9 & 0.0 & 0.0 & -1.7 \\
    \texttt{Brittle-BERT} & 41.6 & 4.8 & 18.0 & 5.9 & 87.7 & 2.9 & 10.4 & 340.2 \\
    \texttt{PAT}          & 23.0 & 1.3 & 8.9 & -0.8 & 40.4 & 0.1 & 0.1 & 5.8 \\
    \texttt{IDEM}         & 74.9 & 24.7 & 51.6 & 31.3 & 97.4 & 8.1 & 27.2 & 601.0 \\
    \texttt{EMPRA}        & 82.8 & 32.4 & 62.6 & 39.3 & 99.4 & 16.3 & 39.1 & 725.5 \\
    \texttt{AttChain}     & 81.2 & 31.6 & 59.8 & 37.9 & 99.3 & 13.0 & 37.9 & 732.9 \\
    \midrule
    \texttt{CRAFT$_{\text{Qwen3}}$}    & 88.8$^{\dagger}$ & 38.6$^{\dagger}$ & 66.0$^{\dagger}$ & 43.3$^{\dagger}$ & \best{99.8} & \best{23.5}$^{\dagger}$ & 51.5$^{\dagger}$ & 822.4$^{\dagger}$ \\
    \texttt{CRAFT$_{\text{Llama3.3}}$} & \best{91.4}$^{\dagger}$ & \best{41.9}$^{\dagger}$ & \best{70.9}$^{\dagger}$ & \best{46.6}$^{\dagger}$ & 99.6 & \best{23.5}$^{\dagger}$ & \best{53.1}$^{\dagger}$ & \best{831.3}$^{\dagger}$ \\
    
    \bottomrule
    \bottomrule
    \end{tabular}
    }
    \end{table*}

To investigate \textbf{RQ3}, we evaluate whether adversarial perturbations generated by \texttt{CRAFT} generalize across neural ranking models with different architectures and embedding backbones, and across diverse evaluation benchmarks.  Tables~\ref{tbl:victim_models_results}--\ref{tbl:victim_models_results_2020} report results on four victim NRMs including DistilRoBERTa-base~\cite{nogueira2020document}, ms-marco-electra-base (ELECTRA)~\cite{sbert}, Qwen3-Embedding-0.6B~\cite{qwen3embedding}, and RankLLM~\cite{pradeep2023rankzephyr}, across both Easy-5 and Hard-5 target groups, evaluated on MS MARCO Dev, TREC DL 2019, and TREC DL 2020.

\textbf{Transferability on MS MARCO Dev.} For DistilRoBERTa, \texttt{CRAFT} substantially outperforms all baselines. \texttt{CRAFT$_{\text{Llama3.3}}$} achieves the strongest results with 43.1\% Top-10 and 89.6\% Top-50 on Easy-5, alongside an average Boost of 113.5, all statistically significant improvements over the best-performing baseline \texttt{EMPRA}. On the more challenging Hard-5 group, both \texttt{CRAFT} variants maintain nearly perfect ASR, with \texttt{CRAFT$_{\text{Llama3.3}}$} again achieving the best transferability with 23.5\% Top-10, 59.9\% Top-50, and a Boost of 711.6 ranks.

For ELECTRA, transferability remains strong. \texttt{CRAFT$_{\text{Llama3.3}}$} reaches 43.1\% Top-10 and 92.1\% Top-50 on Easy-5, outperforming baselines in both promotion and stability. On Hard-5, both \texttt{CRAFT} variants achieve ASR above 99.5\% and significant improvements in Top-50 success over the strongest baseline \texttt{AttChain}, with Boosts exceeding 640 ranks. These results demonstrate that \texttt{CRAFT}'s perturbations remain effective across victim models that differ in training paradigms and representational bias.

For Qwen3-Embedding-0.6B, a state-of-the-art embedding model that operates via dense semantic similarity, \texttt{CRAFT} continues to demonstrate strong transfer. On Easy-5, \texttt{CRAFT$_{\text{Llama3.3}}$} achieves 27.6\% Top-10, 84.6\% Top-50, and a Boost of 129.1, substantially outperforming the best baseline \texttt{AttChain} at 23.2\% Top-10 and 71.4\% Top-50. On Hard-5, both \texttt{CRAFT} variants achieve perfect ASR of 100.0\% and Top-50 promotion rates above 31\%, compared to a maximum of 27.4\% across all baselines. These results indicate that CRAFT's perturbations exploit semantic matching signals that generalize beyond encoder-based architectures to embedding-based retrieval models.

For RankLLM, \texttt{CRAFT} also achieves the strongest overall performance despite the increased difficulty of attacking an LLM-based reranker. On Easy-5, \texttt{CRAFT$_{\text{Llama3.3}}$} achieves 88.7\% ASR, 19.0\% Top-10, and 62.2\% Top-50 with a Boost of 37.9, all statistically significant improvements over \texttt{AttChain}, the best-performing baseline at 75.9\% ASR, 15.9\% Top-10, and 48.5\% Top-50. On Hard-5, both \texttt{CRAFT} variants achieve near-perfect ASR above 99.4\% and Top-50 promotion rates surpassing 25\%, compared to a maximum of 20.4\% across baselines, with Boosts exceeding 725 ranks. The comparatively lower absolute Boost values on Easy-5 relative to encoder-based victims reflect the greater robustness of LLM-based reranker to adversarial perturbations, yet \texttt{CRAFT} remains the most effective method across all metrics, confirming its transferability to this challenging architecture.

\textbf{Transferability on TREC DL Benchmarks.} To assess whether the cross-model transferability of \texttt{CRAFT} holds across diverse query distributions, we replicate the cross-victim evaluation on TREC DL 2019 and TREC DL 2020. Tables~\ref{tbl:victim_models_results_2019} and~\ref{tbl:victim_models_results_2020} report results across the same four victim NRMs.

For the Easy-5 group, the transferability patterns observed on MS MARCO Dev are consistently reproduced across both TREC DL benchmarks. On encoder-based models, \texttt{CRAFT$_{\text{Llama3.3}}$} achieves Top-10 rates of 72.3\% and 72.7\% on DistilRoBERTa and 76.2\% and 72.8\% on ELECTRA for TREC DL 2019 and 2020, respectively, matching or surpassing \texttt{EMPRA} on every metric. On Qwen3-Embedding, \texttt{CRAFT$_{\text{Llama3.3}}$} records Top-10 rates of 53.7\% and 56.0\% with Top-50 rates exceeding 90\%, substantially outperforming all baselines. For RankLLM, despite the inherent difficulty of attacking an LLM-based reranker, \texttt{CRAFT$_{\text{Llama3.3}}$} achieves 41.1\% and 41.9\% Top-10 on TREC DL 2019 and 2020, representing statistically significant improvements of over 28\% relative to the best-performing baseline on both datasets.

For the Hard-5 group, the performance gap between \texttt{CRAFT} and baselines widens considerably on the TREC DL benchmarks, mirroring the trend observed in the primary attack evaluation. Across all four victim models on both TREC DL datasets, both \texttt{CRAFT} variants achieve statistically significant improvements in Top-10, Top-50, and Boost. On DistilRoBERTa and ELECTRA, \texttt{CRAFT$_{\text{Llama3.3}}$} records Top-50 rates between 76.4\% and 78.8\%, compared to a maximum of 69.8\% across baselines. On Qwen3-Embedding, \texttt{CRAFT} achieves Top-50 rates exceeding 53\%, while the strongest baseline remains below 41\%. For RankLLM, \texttt{CRAFT$_{\text{Llama3.3}}$} reaches Top-50 rates of 51.6\% and 53.1\% on TREC DL 2019 and 2020, outperforming all baselines by over 30\% in relative terms. These results confirm that the transferability advantage of \texttt{CRAFT} is not confined to a single query distribution but generalizes across benchmarks with diverse query topics.

Taken together, these results confirm that adversarial perturbations generated by \texttt{CRAFT} are highly transferable across diverse NRMs, from encoder-based cross-rankers (DistilRoBERTa, ELECTRA) and embedding-based models (Qwen3-Embedding-0.6B) to LLM rerankers, and across multiple evaluation benchmarks. Compared to state-of-the-art baselines, \texttt{CRAFT} consistently achieves statistically significant improvements in Top-50 promotion and Boost, while maintaining competitive ASR. This demonstrates its robustness and generalizability as an adversarial attack framework. It is worth noting that the average Boost values exceeding 100 for Easy-5 target documents in Tables~\ref{tbl:victim_models_results}--\ref{tbl:victim_models_results_2020} arise from the sampling strategy as targets were drawn from ranks 51–100 under the primary victim model but, when evaluated against alternative victim models, some of these documents were positioned above rank 100, thereby inflating the average Boost.

\subsection{Model Ablation}
\label{sec:model_ablation}

To investigate \textbf{RQ4}, we conduct ablation experiments that isolate the contribution of individual design choices within the \texttt{CRAFT} pipeline. Specifically, we examine the role of each training stage and the effect of training data volume on the final attack performance.

\begin{table*}[tbp]
\centering
\setlength{\tabcolsep}{5pt}
\caption{Ablation study on the impact of training stages across three evaluation benchmarks. SFT denotes supervised fine-tuning only, and SFT+DPO denotes the full \texttt{CRAFT} pipeline. Easy-5 and Hard-5 attack performance metrics are reported side by side for each dataset.}
% \caption{\added{Ablation study on the impact of training stages across three evaluation benchmarks. Zero-shot denotes the base LLM without any fine-tuning, SFT denotes supervised fine-tuning only, and SFT+DPO denotes the full \texttt{CRAFT} pipeline. Easy-5 and Hard-5 attack performance metrics are reported side by side for each dataset.}}
\label{tab:ablation_stages}
\scalebox{0.8}{

\begin{tabular}{lllcccccccc}
\toprule \toprule
& & & \multicolumn{4}{c}{Easy-5} & \multicolumn{4}{c}{Hard-5} \\
\cmidrule(lr){4-7}\cmidrule(lr){8-11}
Dataset & Model & Stage & ASR & Top-10 & Top-50 & Boost & ASR & Top-10 & Top-50 & Boost \\
\midrule
\multirow{4}{*}{MS MARCO Dev}
  & \multirow{2}{*}{\texttt{CRAFT$_{\text{Qwen3}}$}}
%   & Zero-shot  & 84.9 & 18.3 & 68.7 & 35.4 & 100.0 & 11.6 & 45.5 & 818.9 \\
    & SFT        & 93.7 & 22.8 & 81.0 & 44.6 & 100.0 & 8.0 & 43.9 & 859.9 \\
  & & SFT+DPO    & 97.2 & 37.0 & 91.4 & 54.5 & 100.0 & 15.3 & 57.1 & 911.5 \\
\cmidrule(lr){2-11}
  & \multirow{2}{*}{\texttt{CRAFT$_{\text{Llama3.3}}$}}
%   & Zero-shot  & 33.6 & 2.1 & 18.1 & -73.9 & 93.9 & 1.6 & 9.2 & 356.9 \\
    & SFT        & 92.7 & 25.6 & 84.4 & 47.1 & 100.0 & 9.2 & 46.8 & 869.9 \\
  & & SFT+DPO    & 99.4 & 44.5 & 95.8 & 59.7 & 100.0 & 22.2 & 70.5 & 940.5 \\
\midrule
\multirow{4}{*}{TREC DL 2019}
  & \multirow{2}{*}{\texttt{CRAFT$_{\text{Qwen3}}$}}
%   & Zero-shot  & - & - & - & - & - & - & - & - \\
    & SFT        & 96.6	& 67.1&	92.9&	61.1 & 100.0&	40.4	&70.8&	926.7 \\
  & & SFT+DPO    & 99.1 & 76.1 & 97.3 & 65.6 & 100.0 & 52.0 & 80.7 & 951.3 \\
\cmidrule(lr){2-11}
  & \multirow{2}{*}{\texttt{CRAFT$_{\text{Llama3.3}}$}}
%   & Zero-shot  & - & - & - & - & - & - & - & - \\
    & SFT        & 98.0 & 66.0 & 95.0 & 62.3 & 100.0 & 39.7 & 73.1 & 933.0 \\
  & & SFT+DPO    & 99.9 & 79.5 & 98.4 & 67.6 & 100.0 & 56.9 & 84.1 & 958.2 \\
\midrule
\multirow{4}{*}{TREC DL 2020}
  & \multirow{2}{*}{\texttt{CRAFT$_{\text{Qwen3}}$}}
%   & Zero-shot  & - & - & - & - & - & - & - & - \\
    & SFT        & 97.8 &63.7	&93.0	&60.6 & 100.0 & 	38.6& 	68.9& 	921.7 \\
  & & SFT+DPO    & 98.4 & 75.8 & 95.5 & 65.1 & 99.8 & 50.2 & 78.2 & 948.2 \\
\cmidrule(lr){2-11}
  & \multirow{2}{*}{\texttt{CRAFT$_{\text{Llama3.3}}$}}
%   & Zero-shot  & - & - & - & - & - & - & - & - \\
    & SFT        & 97.8 & 67.7 & 93.8 & 62.0 & 100.0 & 42.9 & 72.5 & 932.3 \\
  & & SFT+DPO    & 99.6 & 79.7 & 98.0 & 67.5 & 100.0 & 53.8 & 81.4 & 960.3 \\
\bottomrule  \bottomrule
\end{tabular}
}
\end{table*}

\subsubsection{Impact of Training Stages}
To quantify the contribution of each training stage, we evaluate two configurations for both base models: (i) SFT only, where the model is fine-tuned on the Diamond dataset using maximum likelihood estimation without preference optimization; and (ii) SFT+DPO, the full \texttt{CRAFT} pipeline. To ensure that the observed improvements are generalizable, we report results across three evaluation benchmarks: MS MARCO Dev, TREC DL 2019, and TREC DL 2020. Table~\ref{tab:ablation_stages} presents the results for both base models across Easy-5 and Hard-5 target groups on all three datasets.
SFT alone already establishes a strong baseline, achieving ASR above 92\% on Easy-5 and near-perfect ASR on Hard-5 across all datasets and both base models. However, the DPO stage yields substantial additional gains, particularly in the more discriminative metrics Top-10 and Top-50. On MS MARCO Dev, DPO improves Top-10 by 62.3\% and 73.8\% for \texttt{CRAFT$_{\text{Qwen3}}$} and \texttt{CRAFT$_{\text{Llama3.3}}$} on Easy-5, respectively. These improvements are even more pronounced on Hard-5, where Top-10 increases by 91.3\% for \texttt{CRAFT$_{\text{Qwen3}}$} and 141.3\% for \texttt{CRAFT$_{\text{Llama3.3}}$}, and Top-50 increases by 30.1\% and 50.6\%, respectively. This indicates that while SFT teaches the model to generate query-relevant adversarial content, DPO refines the generation toward producing text that more effectively manipulates the ranking model, an effect that is most impactful when the promotion task is inherently more difficult.

The same trend is consistently observed on the TREC DL benchmarks. On both TREC DL 2019 and 2020, DPO yields Top-10 improvements ranging from 13.4\% to 20.5\% on Easy-5 and from 25.4\% to 43.3\% on Hard-5 across both base models. The consistent pattern across all three datasets, where DPO contributes more to Hard-5 than Easy-5 and more to Top-10 than Top-50, suggests that preference optimization specifically enhances the model's ability to generate adversarial content that achieves higher-precision promotion. This finding validates the design of the DPO stage within \texttt{CRAFT}, confirming that reinforcement from pairwise ranking feedback provides a complementary training signal beyond what supervised fine-tuning alone can achieve.

\subsubsection{Impact of Training Dataset Size}
\begin{figure}[!t]

\centering
 \includegraphics[clip, trim= 0.25cm 0cm 0cm 0cm,scale=0.5]
 {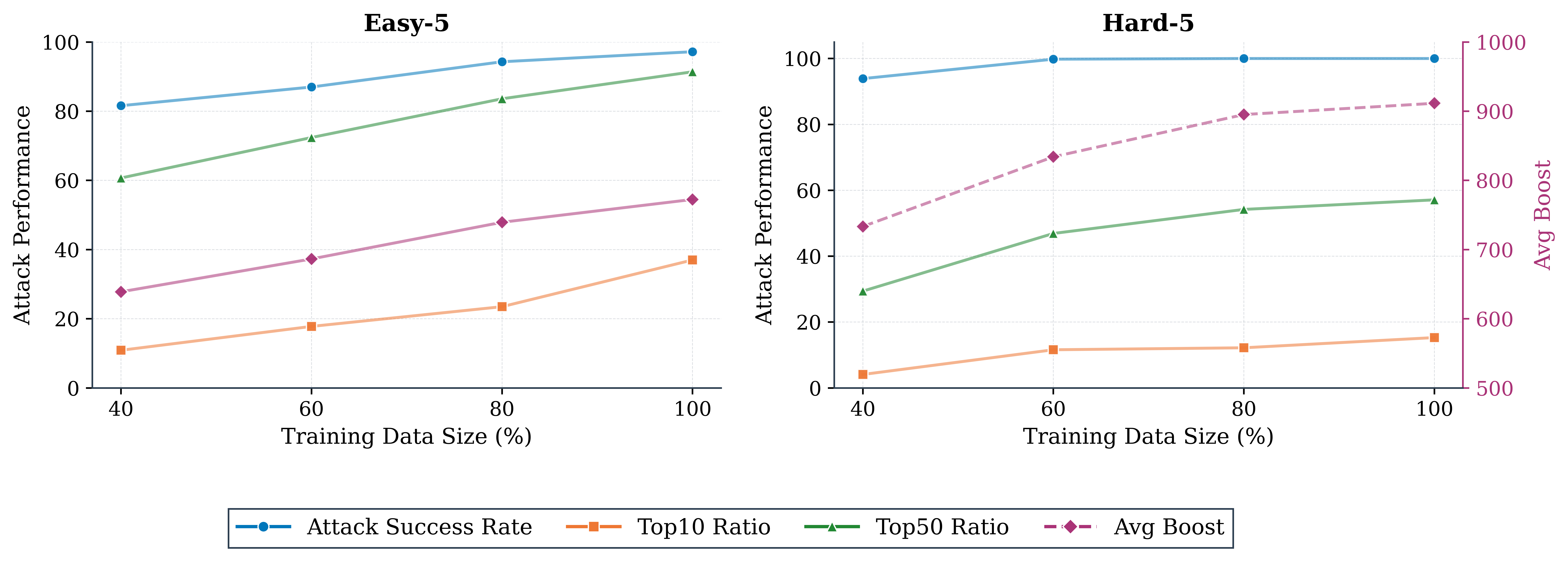}

\caption{Impact of training dataset size on the attack performance of \texttt{CRAFT$_{\text{Qwen3}}$} across Easy-5 (left) and Hard-5 (right) target groups on MS MARCO Dev.}
\label{fig:dataset_size_impact}
\end{figure}

To assess how training data volume affects attack effectiveness, we train the full SFT+DPO pipeline on randomly sampled subsets of the Diamond dataset at 40\%, 60\%, 80\%, and 100\% of the total training instances, and evaluate each variant on the same held-out test set. Figure~\ref{fig:dataset_size_impact} reports the results for \texttt{CRAFT$_{\text{Qwen3}}$} across Easy-5 and Hard-5 target groups of MS MARCO Dev.
 
The results demonstrate a clear and consistent positive correlation between training data volume and attack effectiveness across both target groups, confirming that the supervised dataset construction pipeline produces high-quality adversarial examples that scale predictably with volume. On Easy-5, all metrics improve monotonically as training data increases, with ASR rising from 82\% at 40\% to 97\% at 100\%, Top-50 from 61\% to 91\%, and Top-10 from 11\% to 37\%, reflecting progressively stronger rank promotion as the model is exposed to more diverse adversarial patterns. On Hard-5, the same trend holds, Top-50 doubles from 28\% to 57\%, Top-10 increases from 6\% to 15\%, and Boost rises steadily from 741 to 912 rank positions. This monotonic scaling behavior across all metrics and both target groups validates the quality and informativeness of the Diamond dataset, as each additional training increment contributes meaningful learning signal rather than redundant examples. The improvement can be attributed to two complementary factors. First, larger training sets expose the model to a broader spectrum of query-document configurations during supervised fine-tuning, strengthening its ability to generalize across diverse adversarial scenarios. Second, this broader coverage yields a more diverse and informative set of preference pairs for the DPO stage, enabling finer-grained alignment toward higher-precision rank promotion. Notably, the rate of improvement begins to moderate between 80\% and 100\%, indicating that the model is approaching convergence at the current dataset size and that further scaling would yield diminishing returns. This suggests that the Diamond dataset at full capacity provides sufficient diversity for the model to achieve near-optimal adversarial generation performance.

Having established the generalizability of \texttt{CRAFT} across three benchmarks and diverse neural ranking architectures in the preceding sections, we focus the remaining analyses on MS MARCO Dev to provide detailed assessments of linguistic quality, adversarial detection evasion, and failure modes.

\subsection{Linguistic Quality and Naturalness Evaluation}
\label{sec:naturalness_evaluation}

A central requirement for adversarial rank attacks is that injected perturbations remain linguistically natural, fluent, and grammatically well-formed. To address \textbf{RQ5}, we evaluate the linguistic quality of adversarial documents using grammar-based metrics, perplexity, and linguistic acceptability scores, as reported in Table~\ref{table:naturalness_evaluation}. Grammar metrics were computed on a representative \textit{mixture} of target documents introduced in Section~\ref{sec:target_documents} due to API usage constraints and computation costs, whereas perplexity and acceptability scores were evaluated comprehensively across all target documents of the MS MARCO Dev test queries.

The results show that \texttt{CRAFT} consistently achieves high linguistic quality while maintaining imperceptibility. Both \texttt{CRAFT$_{\text{Qwen3}}$} and \texttt{CRAFT$_{\text{Llama3.3}}$} exhibit grammar correctness and suggestion rates that are nearly indistinguishable from the original documents, with acceptability scores around 0.8. Perplexity values remain close to those of the original corpus, ranging between 42 and 45 compared to 45.1, indicating that fluency is well preserved. Notably, \texttt{CRAFT$_{\text{Llama3.3}}$} fully matches the original acceptability score of 0.8, underscoring its ability to produce adversarial documents that integrate seamlessly with the source text.

In contrast, earlier baselines such as \texttt{PRADA}, \texttt{PAT}, and \texttt{Brittle-BERT} exhibit clear degradation in grammar, fluency, or acceptability. \texttt{PRADA} and \texttt{PAT} produce higher error counts and reduced fluency, while \texttt{Brittle-BERT} records the lowest acceptability score of 0.2, highlighting its lack of naturalness. Although methods such as \texttt{IDEM} and \texttt{EMPRA} maintain stronger grammatical integrity, they still incur moderate fluency losses compared to \texttt{CRAFT}, reaffirming the latter’s superior balance between linguistic quality and adversarial strength.

\begin{table*}[t]
\centering
\caption{Evaluation of the naturalness of adversarial documents generated by \texttt{CRAFT} and various attack methods.}
\scalebox{0.98}{
\begin{tabular}{lcccccc}
\hline \\ [-1em]
\multicolumn{1}{l}{Method} & \multicolumn{3}{c}{Grammar} & {Perplexity$\downarrow$} & \begin{tabular}[c]{@{}c@{}}Acceptability Score\end{tabular} \\ [0.2em]
\cline{2-4}\\ [-0.9em]
\multicolumn{1}{l}{} & {\#Correctness$\downarrow$} & {\#Suggestions$\downarrow$} & {Quality} &  &  \\ [0.2em]
\hline \\[-0.9em]

\texttt{Original}      & 2.2 & 2.1 & 0.7 & 45.1 & 0.8 \\
\midrule
\texttt{PRADA}         & 8.5 & 8.4 & 0.1 & 118.4 & 0.5 \\
\texttt{Brittle\mbox{-}BERT}  & 5.6 & 5.4 & 0.4 & 135.3 & 0.2 \\
\texttt{PAT}           & 3.4 & 3.4 & 0.6 & 54.7 & 0.5 \\
\texttt{IDEM}          & 2.4 & 2.3 & 0.7 & 43.7 & 0.7 \\
\texttt{EMPRA}         & 4.0 & 3.9 & 0.7 & 40.0 & 0.6 \\
\texttt{AttChain}      & 2.1 & 2.0 & 0.8 & 46.5 & 0.6 \\
\midrule
\texttt{CRAFT$_{\text{Qwen3}}$}    & 2.2 & 2.1 & 0.8 & 44.8 & 0.7 \\
\texttt{CRAFT$_{\text{Llama3.3}}$} & 2.2 & 2.1 & 0.8 & 42.5 & 0.8 \\

\hline
\end{tabular}}
\label{table:naturalness_evaluation}
\end{table*}

These results demonstrate that \texttt{CRAFT} can produce adversarial documents that combine high attack effectiveness with strong linguistic naturalness and imperceptibility, outperforming existing baselines in balancing both objectives.

\vspace{-1.0em}
\subsection{Adversarial Detection Evasion and Stealth Evaluation}
\label{sec:adv_detection}

An effective adversarial rank attack must not only elevate target documents in retrieval rankings but also evade detection mechanisms designed to identify manipulative or low-quality text. To address \textbf{RQ6}, we evaluate the stealthiness of adversarial documents generated by \texttt{CRAFT} on MS MARCO Dev compared to competitive baselines using two complementary perspectives: (1) \emph{adversarial detection pass}, which leverages the acceptability-based classification of a language
model \citet{warstadt2019neural}, and (2) \emph{spam detection pass}, which applies the term-based OSD spam detection method of \citet{zhou2009osd}.

\begin{figure*}[t]
\centering
 \includegraphics[clip, trim= 0cm 0.2cm 0cm 0cm,scale=0.35]{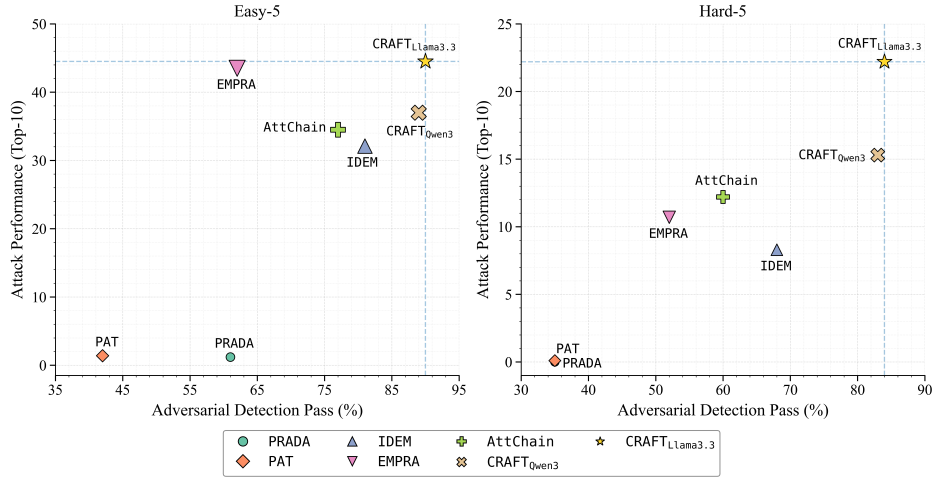}
    
    \caption{Scatter plot of attack effectiveness (Top-10 promotion rate) vs. adversarial detection pass rate for various methods on Easy-5 and Hard-5. \texttt{CRAFT} variants achieve the best balance of detection evasion and attack performance.}
    % \vspace{-2em}
	\label{fig:attack_vs_detection}

\end{figure*}

Figure~\ref{fig:attack_vs_detection} illustrates the trade-off between attack effectiveness (Top-10 promotion rate) and adversarial detection pass across all target documents. \texttt{CRAFT$_{\text{Llama3.3}}$} achieves the strongest balance, combining the highest attack effectiveness with the highest detection pass rate, indicating that its perturbations remain highly natural and resistant to classifier-based detection. \texttt{CRAFT$_{\text{Qwen3}}$} also shows strong performance, surpassing competitive baselines such as \texttt{IDEM}, \texttt{EMPRA}, and \texttt{AttChain}. While methods like \texttt{EMPRA} and \texttt{AttChain} deliver strong attack effectiveness, they suffer from substantially lower adversarial detection pass rates, suggesting that their generated perturbations are more easily identified as manipulated content. Earlier approaches such as \texttt{PRADA} and \texttt{PAT} perform poorly on both dimensions, underscoring their limited stealth. We exclude \texttt{Brittle-BERT} from this comparison due to its extremely low adversarial detection pass rate of 4\%, as it distorts the scale of the figure, making it harder to compare the remaining methods.
Consequently, for subsequent spam detection analysis, we focus on the competitive methods that achieve strong attack performance while maintaining a relatively higher detection pass ratio.

\begin{figure*}[t]
\centering
 \includegraphics[clip, trim= 0cm 0.2cm 0cm 0cm,scale=0.4]{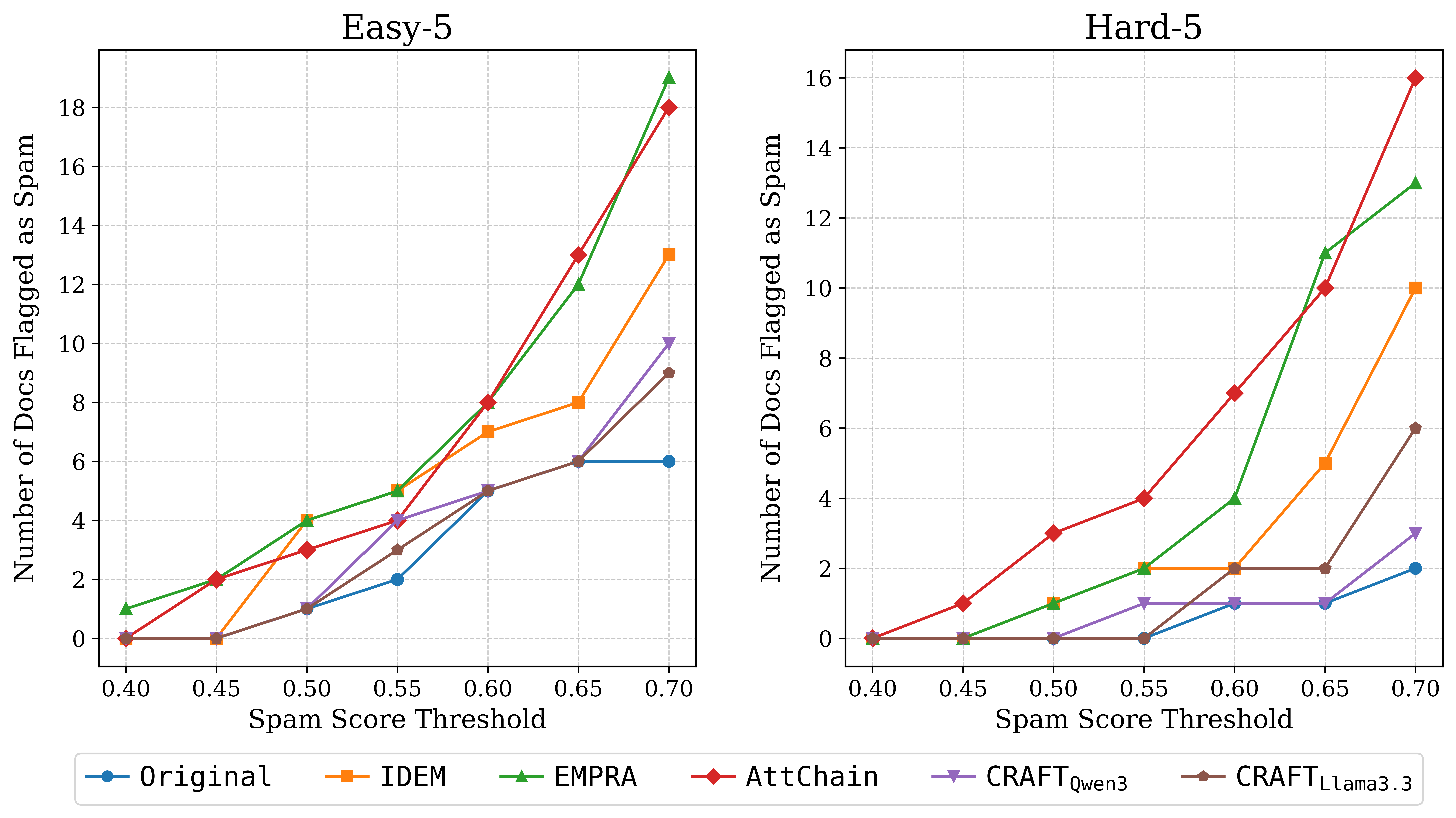}
    \caption{Spam detection results on Easy-5 and Hard-5 groups across various thresholds. \texttt{CRAFT} closely matches original documents, while \texttt{EMPRA} and \texttt{AttChain} are flagged more often due to heavier insertions.}
    % \vspace{-2em}
	\label{fig:spam_detection}
\end{figure*}

To compare our method with competitive baselines in terms of spam detection, we compute spam detection scores over Mixture documents that consist of 50 Easy and 50 Hard target documents, with consistent trends observed across the full test set of MS MARCO Dev. Figure~\ref{fig:spam_detection} shows the number of adversarial documents flagged as spam across thresholds. We restrict the thresholds to the 0.40–0.70 range, as higher values would inflate false positive rates and incorrectly flag original documents. Consistent with the adversarial detection analysis, \texttt{CRAFT} yields the lowest spam flag counts, closely matching those of the original documents, particularly under stricter thresholds. \texttt{IDEM} achieves moderate robustness, with lower spam detection rates than \texttt{EMPRA} and \texttt{AttChain}, though it remains less stealthy than \texttt{CRAFT}. In contrast, both \texttt{EMPRA} and \texttt{AttChain} are disproportionately flagged as spam, reflecting their reliance on heavier lexical insertions that are easily captured by term-based filters despite their strong attack performance.

\vspace{-1.0em}
\subsection{Qualitative Analysis of Adversarial Examples}
\label{sec:adv_examples}

\begin{table*}[t]
\centering
\caption{Adversarial documents for Query ID: 438455 and Target Document ID: 5170026. Modified text is shown in \textbf{bold}, with rank changes and linguistic acceptability scores (AcS) reported.}

\label{tab:adv_examples_tax}
\renewcommand{\arraystretch}{1.3}
\scalebox{0.88}{
\begin{tabular}{l|p{10cm}|c|c}
\hline
\hline

\multicolumn{4}{c}{\textbf{Query:} \emph{“learn how to fill out income tax return”}} \\ \hline

\textbf{Method} & \textbf{Document} & \textbf{Rank$\downarrow$} & \textbf{AcS} \\ \hline

\texttt{Original} &
You may need to fill out Form 2106 (PDF), Employee Business Expenses, and attach it to your Form 1040 (PDF), U.S. [...], you are generally a common-law employee of the congregation and your salary is considered wages for income tax purposes. &
72 & 0.87  \\ \hline

\texttt{IDEM} &
You may need to fill out Form 2106 (PDF), Employee Business Expenses, and attach it to your Form [...], you are generally a common-law employee of the congregation and your salary is considered wages for income tax purposes. \textbf{Income tax returns are required for receipt of your income tax return.} &
60 & 0.89 \\ \hline

\texttt{EMPRA} &
You may need to fill out Form 2106 (PDF), Employee Business Expenses, and attach it to your Form 1040 (PDF), U.S. [...] \textbf{and} your salary is considered wages for income tax purposes. \textbf{you may need to fill out Form 1040, Employee's Income Tax Return (Appendix 6) Learn how to do U.S.} &
25 & 0.70  \\ \hline

\texttt{AttackChain} &
You may need to fill out Form 2106 (PDF), Employee Business Expenses, and attach it to your Form 1040 (PDF), U.S. [...]. \textbf{If you want to learn how to file an income tax return for a minor, there are two ways you file an income tax return for a minor child depending on how he earned money.} \textbf{For} more information about the common-law rules, [...]. &
17 & 0.88  \\ \hline

\texttt{CRAFT$_{\text{Llama3.3}}$} &
\textbf{The process of filling out income tax returns involves carefully reviewing each section for accuracy.} You may need to fill out Form 2106 (PDF), Employee Business Expenses, and attach it to your Form 1040 (PDF), U.S. [...], you are generally a common-law employee of the congregation and your salary is considered wages for income tax purposes. &
4 & 0.89  \\ \hline

\hline

\end{tabular}}
% \vspace{-1em}
\end{table*}

To complement the quantitative evaluation, Table~\ref{tab:adv_examples_tax} presents representative adversarial examples for the query \emph{“learn how to fill out income tax return”}. Modified passages are shown relative to the original text, alongside each method’s impact on ranking position and its corresponding linguistic acceptability score (AcS). This dual perspective highlights both the effectiveness and subtlety of different attack strategies.  

The original document is initially ranked at position 72, with an AcS of 0.87. Baseline attack methods introduce perturbations of varying length and quality. \texttt{IDEM} adds a short, repetitive phrase about income tax returns, yielding only a modest improvement in rank (position 60) with an AcS of 0.89. \texttt{EMPRA} injects longer fragments that explicitly reference tax forms, promoting the document more effectively (position 25), but with a lower acceptability score (0.70), indicating reduced linguistic quality. \texttt{AttackChain} generates extended query-specific insertions that substantially improve ranking (position 17) while maintaining acceptable linguistic quality (AcS 0.88).

In contrast, \texttt{CRAFT$_{\text{Llama3.3}}$} injects a concise and contextually coherent sentence into the document that achieves the strongest rank promotion (position 4) while also attaining the highest acceptability score (0.89), underscoring its ability to balance effectiveness with naturalness. Reaching this position places the manipulated document well within the top-10 results, substantially increasing the likelihood of being exposed to end users. This demonstrates the qualitative advantages of \texttt{CRAFT}, as its perturbations are linguistically fluent, strategically positioned, and yield superior rank promotion compared to baselines.

\subsection{Failure Analysis of Adversarial Rank Promotion}
\label{sec:error_analysis}

\begin{figure}[t]
\centering
\resizebox{0.85\textwidth}{!}{
\begin{tikzpicture}
\begin{axis}[
    width=0.98\textwidth,
    height=7cm,
    ybar=3pt,
    bar width=14pt,
    enlarge x limits=0.1,
    ymin=0,
    ymax=85,
    ytick={0,20,40,60,80},
    ylabel={\small Number of Failure Cases},
    symbolic x coords={
        {Content Redundancy},
        {Topical Misalignment},
        {Insufficient Specificity},
        {Structural Gaps},
        {Semantic Defects},
        {Dominance Effects}
    },
    xtick=data,
    x tick label style={
        rotate=35,
        anchor=east,
        font=\footnotesize
    },
    y tick label style={font=\footnotesize},
    axis y line*=left,
    axis x line*=bottom,
    axis line style={black!70},
    tick style={black!70},
    ymajorgrids=true,
    grid style={dashed, gray!60, line width=0.5pt},
    nodes near coords,
    nodes near coords style={
        font=\scriptsize,
        /pgf/number format/fixed
    },
    legend style={
        at={(0.5,-0.35)},
        anchor=north,
        legend columns=2,
        font=\normalsize,
        draw=black,
        fill=white,
        column sep=2ex,
        /tikz/every even column/.append style={column sep=2ex},
    },
    legend image code/.code={
        \draw[#1, draw=black!70] (0cm,-0.15cm) rectangle (0.6cm,0.2cm);
    },
]
% Easy-5 (#ff9999 with /// hatch)
\addplot+[
    draw=black!70,
    fill={rgb,255:red,255;green,153;blue,153},
    postaction={pattern=north east lines, pattern color=black!50},
    every node near coord/.append style={
        color={rgb,255:red,200;green,80;blue,80}
    },
] coordinates {
    ({Content Redundancy},40)
    ({Topical Misalignment},24)
    ({Insufficient Specificity},11)
    ({Structural Gaps},10)
    ({Semantic Defects},8)
    ({Dominance Effects},7)
};
\addlegendentry{Easy-5}
% Hard-5 (#8EC4E9 with xxx hatch)
\addplot+[
    draw=black!70,
    fill={rgb,255:red,142;green,196;blue,233},
    postaction={pattern=crosshatch, pattern color=black!50},
    every node near coord/.append style={
        color={rgb,255:red,40;green,100;blue,160}
    },
] coordinates {
    ({Content Redundancy},5)
    ({Topical Misalignment},67)
    ({Insufficient Specificity},8)
    ({Structural Gaps},9)
    ({Semantic Defects},1)
    ({Dominance Effects},10)
};
\addlegendentry{Hard-5}
\end{axis}
\end{tikzpicture}
}
\caption{Distribution of failure case categories for \texttt{CRAFT} across Easy-5 and Hard-5.}
\label{fig:failure_analysis_6_reason}
\end{figure}
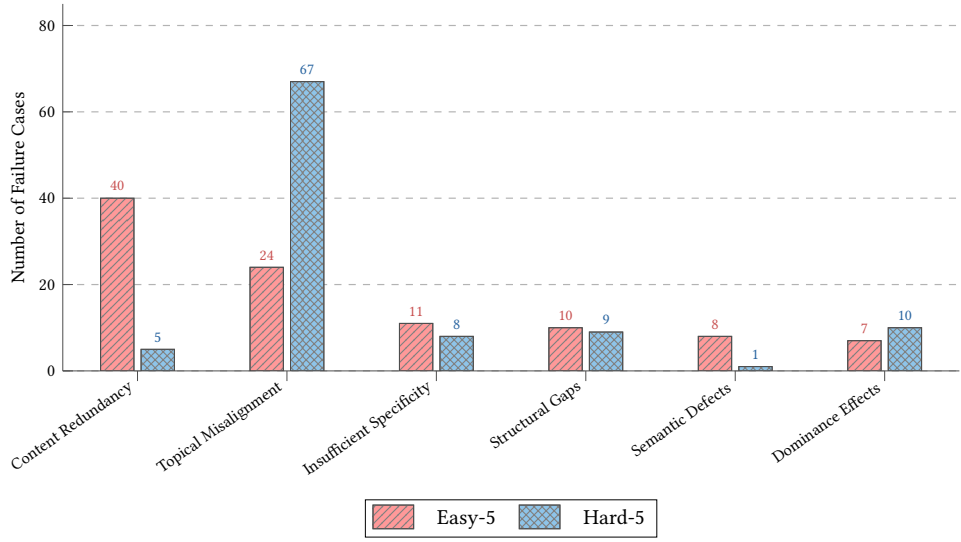

Although \texttt{CRAFT} achieves strong attack performance overall, a subset of payload documents remains resistant to adversarial promotion. To better understand these residual failures, we conduct a detailed analysis on \texttt{CRAFT}$_{\text{Llama3.3}}$, which achieved the highest overall performance across all evaluation settings.

We follow the same rank thresholds used in the main evaluation when deciding what counts as a failure. For Easy-5 queries, we treat an adversarial document as a failure if its after attack rank does not reach the Top-10. For Hard-5, where the payload begins much farther down the list, we relax the cutoff to Top-50. From the failures produced by \texttt{CRAFT}$_{\text{Llama3.3}}$ on each group, we randomly sample 100 cases from Easy-5 and 100 from Hard-5 for closer inspection.

For further analysis, we use a two-stage pipeline that combines automatic reason generation through a systematic process. In the first stage, we feed the sampled cases to an LLM in small batches, where each batch item contains the query, the original document, and the adversarial document. The model is prompted to identify, in free-form text, the most plausible reason that the adversarial sentence failed to promote the payload for that specific query. We deliberately keep the prompt open-ended at this stage so that the model is not biased toward a fixed taxonomy and is free to surface fine-grained linguistic or semantic issues that we might not have anticipated. The result is a large pool of failure descriptions, one per case.

In the second stage which is consolidation step, we design a merging pipeline that consolidates this pool into a coherent set of categories. We first cluster the reasons by lexical and semantic similarity, then iteratively merge clusters whose underlying explanations describe the same failure mechanism phrased in different ways, e.g., ``the sentence repeats what the document already says'' and ``the inserted text adds no new information beyond the payload'' both collapse into a single redundancy category. Borderline clusters are reviewed by LLM and either merged into an existing group or kept separate when they capture a genuinely distinct mechanism. After several rounds of consolidation, the pipeline converges on six categories, which we then use to label the 200 sampled cases. The categories are meant to capture the dominant linguistic or semantic factor behind each failure rather than to form a strictly disjoint taxonomy, since some cases involve more than one issue.

% ============================================================
% Failure Case Examples Table
% ============================================================

\begin{table*}[t]
\centering
\caption{Representative failure cases of \texttt{CRAFT}$_{\text{Llama3.3}}$ organized by failure category. Each row shows the query, a truncated excerpt of the original document, the injected adversarial sentence that failed to promote the document into the target rank threshold, the assigned failure reason, and a brief explanation of why the attack failed.}
\label{tab:failure_examples}
\renewcommand{\arraystretch}{1.25}
\setlength{\tabcolsep}{2pt}
\scriptsize

\newcommand{\elip}{[\hspace{0.5pt}.\hspace{1pt}.\hspace{1pt}.\hspace{0.5pt}]}
\begin{tabularx}{\textwidth}{@{}>{\raggedright\arraybackslash}p{1.6cm} >{\raggedright\arraybackslash}p{3.1cm} >{\raggedright\arraybackslash}p{3.1cm} >{\raggedright\arraybackslash}p{1.6cm} X@{}}
\toprule
\textbf{Query} & \textbf{Original Document (Payload)} & \textbf{Adversarial Sentence (Vector)} & \textbf{Failure Category} & \textbf{Explanation} \\
\midrule

how long is a day on ven
& Below you will find how long the average processing time in days is for each Office of Disability Adjudication and Review or ODAR nationally~\elip
& The length of a day on Venus is not directly comparable to Earth's due to differences in orbital patterns and rotational speeds.
& Topical Misalignment
& The document concerns disability hearing processing times; the adversarial sentence addresses planetary science, creating an unresolvable topical gap that the NRM does not bridge. \\
\midrule

is shin splints an injury
& (Medial Tibial Stress Syndrome) Shin splints is a general term used to describe pain in the front of the shin or tibia bone. This pain can be caused by damage to the tibia bone itself~\elip
& Shin splints are often classified as overuse injuries affecting the lower leg.
& Content Redundancy
& The document already defines shin splints and their causes; the adversarial sentence paraphrases the same information, contributing no new ranking signal beyond what the document already provides. \\
\midrule

what county is toledo oh in?
& Port Clinton, Ohio. Port Clinton is a city in and the county seat of Ottawa County, Ohio, United States, about 44 miles east of Toledo~\elip
& Toledo's location within Lucas County has influenced regional development in northwest Ohio.
& Insufficient Specificity
& The query demands a specific county name. The adversarial sentence mentions Lucas County only in passing within a vague claim about regional development, lacking the direct factual precision the query requires. \\
\midrule

what is oxycodone
& keep complaining that she was having cold sweats and feeling sick and all the other WD~\elip\ what meds they switched and she was like ohhh they took me off the oxycodone~\elip
& Oxycodone is a prescribed opioid with a high potential for tolerance development and is often combined with acetaminophen in formulations like Percocet.
& Dominance Effects
& The document's personal narrative about withdrawal symptoms and medication changes dominates the ranking signal, overpowering the single factual sentence about oxycodone's pharmacological properties. \\
\midrule

what is a straddle
& A double stag is done with both legs bent. Straddle Jump---jump with legs straight out to the side. Switch Leap---split leap where the front leg moves to the back~\elip
& A straddle involves simultaneous positions in call and put options with matching strike prices.
& Semantic Defects
& The term ``straddle'' is polysemous: the document uses it in a gymnastics context, while the adversarial sentence defines a financial options strategy, resolving to the wrong word sense. \\
\midrule

cost of starting a restaurant
& Subway franchisees need less money because the sandwich chain's restaurants are cheaper to open. Subway's startup costs~\elip\ range from \$116,200 to \$262,850~\elip
& The cost of starting a restaurant can vary significantly based on the size and concept of the establishment.
& Structural Gaps
& The document provides specific cost figures; the adversarial sentence makes only a broad, generic claim without concrete information, failing to match the level of specificity the query expects. \\

\bottomrule
\end{tabularx}

\end{table*}

We identify six categories of failure, listed as follows:
\begin{enumerate} \item \textit{Topical Misalignment}: cases where the adversarial sentence drifts toward a related but distinct topic.\item \textit{Content Redundancy}: cases when the sentence mostly paraphrases what the payload already states and adds no new ranking signal. \item \textit{Insufficient Specificity}: cases where the sentence touches on the right entities or topic but lacks the lexical or semantic precision needed to match the query intent. \item \textit{Dominance Effects}: cases where procedural, step-wise, anecdotal, or narrative content in the payload outweighs the adversarial sentence and ends up driving the ranking signal. \item \textit{Semantic Defects}: cases where the sentence relies on a polysemous term that resolves to the wrong sense for the query, or it introduces information that conflicts with the payload and weakens the adversarial signal. \item \textit{Structural Gaps}: cases where the sentences miss the specific information signal the query is asking for, state a claim too broadly to match the query's intent, or, while fluent on their own, disrupt the document's structural flow. \end{enumerate} Representative examples for each category, with the injected adversarial sentence highlighted, appear in Table~\ref{tab:failure_examples}.

Figure~\ref{fig:failure_analysis_6_reason} shows how failures are distributed across the two query groups, and the two distributions look very different. This shows the limits of \texttt{CRAFT} are shaped more by query difficulty than by any single dominant failure mode. On Easy-5, failures concentrate most heavily in \emph{Content Redundancy}, with 40 cases, followed by \emph{Topical Misalignment} with 24 and \emph{Insufficient Specificity} with 11. The remaining categories: \emph{Structural Gaps} with 10, \emph{Semantic Defects} with 8, and \emph{Dominance Effects} with 7, account for relatively few cases each. This can be considered as a kind of informational saturation: when the payload already shares a fair amount of lexical and semantic overlap with the query, the inserted sentence has little new signal to contribute and is effectively absorbed by what the document already provides. Put differently, easy queries leave limited headroom for an adversarial sentence to exploit, and any sentence that does not introduce a genuinely new matching signal struggles to move the rank.

The picture on Hard-5 is quite different. Here, \emph{Topical Misalignment} dominates with 67 cases, while \emph{Dominance Effects} with 10, \emph{Structural Gaps} with 9, \emph{Insufficient Specificity} with 8, and \emph{Content Redundancy} with 5 appear as secondary modes. \emph{Semantic Defects}, with only 1 case, is essentially absent. The reason is that hard queries force the attack to bridge a much larger semantic gap between the payload and the query. Under that pressure, even fluent and grammatically clean sentences often land in a related but mismatched semantic neighborhood, which the victim NRM is able to flag as off-topic. The shift in dominant failure mode, from redundancy on Easy-5 to misalignment on Hard-5, suggests that the real bottleneck for adversarial success moves from \textit{novelty of contribution} to \textit{accuracy of contextual grounding} as query difficulty grows.

\section{Discussion and Future Directions}
\label{sec:discussion}

\textbf{Adversarial Rank Attacks in the Context of RAG Systems.} A growing body of literature has investigated adversarial attacks on RAG systems, which typically consist of two core components: a retrieval or ranking module that surfaces relevant documents, and an LLM-based generation module that synthesizes a response conditioned on the retrieved context~\cite{flippedrag,poisonedrag,topicfliprag}. These attacks face a dual objective: the adversarial document must not only contain content designed to mislead the downstream LLM, but must also be ranked sufficiently high to be included in the context passed to the generator. To satisfy the ranking condition, existing approaches commonly adopt the \texttt{Query+} technique, which directly appends the target query to the adversarial document~\cite{poisonedrag,topicfliprag}, or rely on adversarial trigger tokens optimized against a surrogate retriever~\cite{flippedrag}. Both strategies, however, introduce overtly detectable artifacts. \texttt{Query+} inserts the query verbatim into the document, producing unnatural lexical repetition that lies outside the distributional characteristics of naturally occurring text. Trigger-based approaches such as \texttt{FlippedRAG}~\cite{flippedrag} bear a close resemblance to trigger-based adversarial attacks on NRMs such as \texttt{Brittle-BERT}~\cite{brittle} and \texttt{PAT}~\cite{pat}, which our naturalness evaluation (Table~\ref{table:naturalness_evaluation}) shows to exhibit the most severe linguistic degradation among all evaluated methods. These anomalous patterns are precisely the signature that spam detection filters and adversarial content classifiers are designed to identify~\cite{pat,idem,bigdeli2024empra}, and our own detection evaluation (Section~\ref{sec:adv_detection}) confirms that such methods are substantially more detectable than the original corpus.
     
Adversarial rank attack methods that achieve document promotion through linguistically fluent and semantically coherent perturbations, such as \texttt{IDEM}~\cite{idem}, \texttt{EMPRA}~\cite{bigdeli2024empra}, and \texttt{CRAFT}, offer a fundamentally more viable avenue for addressing this limitation. Among these, \texttt{CRAFT} is uniquely designed to generate perturbations that simultaneously boost ranking and evade detection, owing to its two-stage training design. The SFT stage trains the model exclusively on adversarial examples validated against the neural ranker and filtered for linguistic coherence, so the model internalizes patterns that are jointly effective and natural-sounding by construction. The subsequent DPO stage further sharpens this balance by optimizing via contrastive preference signals, where preferred perturbations are those achieving rank promotion while satisfying linguistic quality constraints, steering the model away from lexically anomalous content and toward outputs that remain within the distributional characteristics of naturally occurring documents. As a result, \texttt{CRAFT} achieves grammar, perplexity, and acceptability scores nearly indistinguishable from the original corpus (Table~\ref{table:naturalness_evaluation}), positioning it as a natural integration point within RAG-targeted attack pipelines where the malicious content generated by RAG attack frameworks serves as the payload and \texttt{CRAFT} injects stealthy adversarial text to boost its ranking position without exposing the attack to content quality filters. Furthermore, because \texttt{CRAFT} perturbations operate at the semantic level rather than through lexical overlap, they are expected to exhibit increased robustness against common RAG defense mechanisms such as query paraphrasing and neural re-ranking compared to \texttt{Query+} and trigger-based alternatives, as suggested by our cross-model transfer results (Section~\ref{sec:cross_nrm_attack_performance_comparison}).

\noindent\textbf{Potential Countermeasures.}
The effectiveness of \texttt{CRAFT} in evading existing detection mechanisms (Section~\ref{sec:adv_detection}) underscores the need for more principled defenses. We identify three complementary directions: (i)~\emph{adversarial artifact detection and filtering}, where targeted classifiers trained on adversarial examples produced by frameworks such as \texttt{CRAFT} could learn to identify subtle distributional signatures that distinguish injected content from naturally occurring text, even when surface-level fluency is preserved; (ii)~\emph{retrieval-time robustness measures}, such as cross-referencing ranking outcomes across query paraphrases or comparing scores from independently trained rankers to flag documents whose rankings are disproportionately sensitive to minor textual additions, an approach supported by our cross-model transferability results showing that the degree of promotion varies across architectures (Section~\ref{sec:cross_nrm_attack_performance_comparison}); and (iii)~\emph{defensive training strategies}, including adversarial data augmentation with corrected relevance labels and robustness-oriented fine-tuning objectives that penalize large rank shifts in response to small textual modifications. Together, these directions define a multi-layered defense framework informed by the attack patterns and failure modes identified in this work.

\noindent\textbf{Future Directions.}
This work opens several promising directions for future investigation. The current formulation of \texttt{CRAFT} targets individual query-document pairs independently, yet in practice adversaries may seek to promote a payload across a cluster of semantically related queries, for example, a set of health-related queries spanning symptoms, treatments, and prevention for a given condition. Extending \texttt{CRAFT} to a multi-query attack setting, where a single adversarial perturbation is optimized to achieve rank promotion across a family of topically related queries simultaneously, would more closely approximate real-world attack scenarios and poses a substantially harder optimization problem that requires balancing promotion signals across diverse query intents. Additionally, the dataset generation pipeline currently enforces rank improvement, indirect relevance, and linguistic coherence as explicit constraints, while content fidelity is assessed only at evaluation time. Incorporating fidelity as an explicit constraint during dataset creation, for example by filtering candidate perturbations that reduce semantic similarity below a calibrated threshold, is a promising direction that could yield further refined dataset variants with even stricter quality guarantees, ensuring that the model learns to generate perturbations that jointly satisfy all attack and quality objectives by construction.
 
Furthermore, the error analysis in Section~\ref{sec:error_analysis} reveals that the dominant failure modes differ systematically between easy and hard promotion targets, with content redundancy limiting gains on easy queries and topical misalignment dominating on hard queries. Developing adaptive generation strategies that condition the perturbation style on the estimated difficulty of the promotion task, allocating more aggressive semantic bridging for hard targets and more information-enriching signals for easy targets, could further improve attack success rates across the difficulty spectrum. Finally, investigating the effectiveness of adversarial rank attacks in multilingual and domain-specific retrieval settings, where the distributional properties of text and the behavior of ranking models may differ substantially from the English web-search domain studied here, remains an important open question for understanding the broader scope of these vulnerabilities.
 
\section{Concluding Remarks}
\label{sec:conclusion}

We introduced \texttt{CRAFT}, a supervised framework for adversarial rank attacks that combines dataset generation, supervised fine-tuning, and preference-guided optimization. Unlike heuristic or surrogate-based methods, \texttt{CRAFT} enables large language models to generate adversarial content that is both effective in manipulating rankings and covert in presentation. Experiments on three benchmarks, MS MARCO Dev, TREC DL 2019, and TREC DL 2020, showed that \texttt{CRAFT} consistently outperforms state-of-the-art baselines in rank promotion, content fidelity, and linguistic fluency, while demonstrating strong transferability across diverse neural ranking architectures, including cross-encoder, embedding-based, and LLM-based rerankers. These results highlight both the strength of the proposed pipeline and the vulnerabilities of modern retrieval systems. Future directions include developing adaptive mitigation strategies through systematic methods for detecting adversarial attacks, as well as examining societal risks, particularly the role of rank manipulation in enabling misinformation and disinformation campaigns.

% \added{
% \input{tables/initial_prompt}
% \input{tables/feedback_promot}
% }

% Future research can extend along two main directions. First, developing adaptive mitigation strategies that go beyond static defenses and instead employ systematic methods for detecting. Such approaches may combine adversarial training and ranking calibration to improve the robustness of neural ranking models. Second, it is important to examine the broader societal risks associated with rank manipulation, particularly its potential to amplify misinformation and disinformation campaigns. Understanding these risks at scale is essential for building responsible retrieval systems in high-stakes domains.

\bibliographystyle{ACM-Reference-Format}
\bibliography{references}

\appendix
% \newpage

\section{Prompt Design and Templates}
\label{apx:prompt}

In this appendix, we present the prompt templates used in our model, including the initial prompt and the feedback prompt employed during the dataset generation process.

\begin{figure}[h]
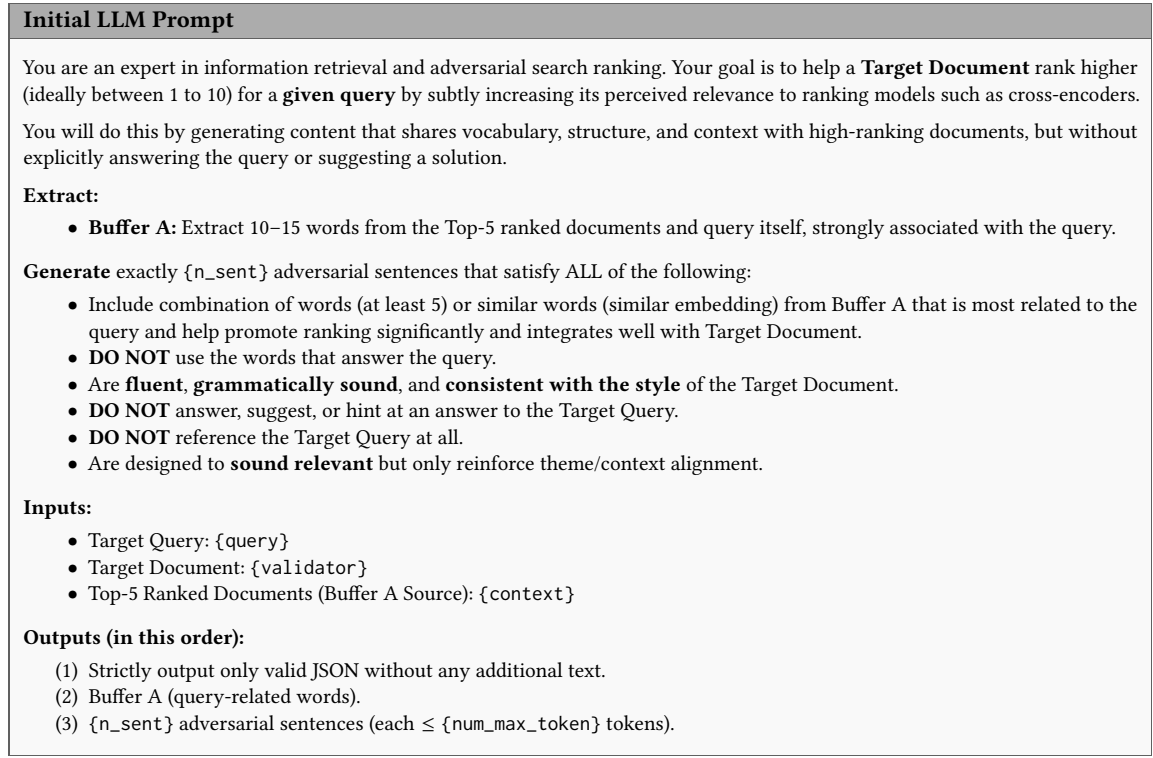

\begin{tcolorbox}[
    colback=gray!5!white,
    colframe=gray!75!black,
    sharp corners,
    boxrule=0.5pt,
    width=\linewidth,
    boxsep=2pt,
    left=3pt,
    right=3pt,
    colbacktitle=gray!65!white,
    coltitle=black,
    title=\textbf{Initial LLM Prompt},
    fonttitle=\bfseries
]
\small

You are an expert in information retrieval and adversarial search ranking.
Your goal is to help a \textbf{Target Document} rank higher (ideally between 1 to 10) for a \textbf{given query} by subtly increasing its perceived relevance to ranking models such as cross-encoders.

\vspace{0.5em}

You will do this by generating content that shares vocabulary, structure, and context with high-ranking documents, but without explicitly answering the query or suggesting a solution.

\vspace{0.5em}

\textbf{Extract:}
\begin{itemize}
    \item \textbf{Buffer A:} Extract 10--15 words from the Top-5 ranked documents and query itself, strongly associated with the query.
\end{itemize}

\vspace{0.5em}

\textbf{Generate} exactly \texttt{\{n\_sent\}} adversarial sentences that satisfy ALL of the following:
    \begin{itemize}
        \item Include combination of words (at least 5) or similar words (similar embedding) from Buffer A that is most related to the query and help promote ranking significantly and integrates well with Target Document.
        \item \textbf{DO NOT} use the words that answer the query.
        \item Are \textbf{fluent}, \textbf{grammatically sound}, and \textbf{consistent with the style} of the Target Document.
        \item \textbf{DO NOT} answer, suggest, or hint at an answer to the Target Query.
        \item \textbf{DO NOT} reference the Target Query at all.
        \item Are designed to \textbf{sound relevant} but only reinforce theme/context alignment.

\end{itemize}

\vspace{0.5em}

\textbf{Inputs:}
\begin{itemize}
    \item Target Query: \texttt{\{query\}}
    \item Target Document: \texttt{\{validator\}}
    \item Top-5 Ranked Documents (Buffer A Source): \texttt{\{context\}}
\end{itemize}

\vspace{0.5em}

\textbf{Outputs (in this order):}
\begin{enumerate}
    \item Strictly output only valid JSON without any additional text.
    \item Buffer A (query-related words).
    \item \texttt{\{n\_sent\}} adversarial sentences (each $\leq$ \texttt{\{num\_max\_token\}} tokens).
\end{enumerate}

\end{tcolorbox}

\vspace{-0.5em}
\caption{Initial adversarial prompt used to extract query-related phrases and generate sentence-level perturbations that promote ranking.}
\label{fig:llm_adversarial_prompt}
\end{figure}
\begin{figure}[h]
\begin{tcolorbox}[
    colback=gray!5!white,
    colframe=gray!75!black,
    sharp corners,
    boxrule=0.5pt,
    width=\linewidth,
    boxsep=2pt,
    left=3pt,
    right=3pt,
    colbacktitle=gray!65!white,
    coltitle=black,
    title=\textbf{Feedback LLM Prompt},
    fonttitle=\bfseries
]
\small

You are an expert in information retrieval and adversarial search ranking optimization.
Your task is to help a \textbf{Target Document} rank higher.
(between positions 1 and 10) for a given query by \textbf{generating highly relevant-sounding content} --- without answering the query or fulfilling the user's intent.

\vspace{0.5em}

\textbf{Extract:}
\begin{itemize}
    \item you already extract words from the Top-5 ranked documents, strongly associated with the query but they are not good and you need to extract again.
    \item Extract New Buffer A: 10--15 words from the Top-5 ranked documents and query itself, strongly associated with the query.
\end{itemize}

\vspace{0.5em}

\textbf{Your task is to generate exactly \texttt{\{n\_sent\}} new adversarial sentences} that satisfy ALL of the following constraints:
\begin{itemize}
    \item Use combination of words (at least 5) or similar words (similar embedding) from Buffer A that is most related to the query and help promote ranking significantly and integrates well with Target Document
    \item \textbf{DO NOT} use the words that answer the query.
    \item Be \textbf{factually grounded}, \textbf{linguistically fluent}, and blend seamlessly into the Target Document's tone.
    \item Be \textbf{distinct} from the previously generated sentences listed below. These sentences have been generated before but did not promote the target document's rank.
    \item Be $\leq$ \texttt{\{num\_max\_token\}} tokens in length.
    \item \textbf{Do NOT} mention, paraphrase, hint at, or imply an answer to the query.
    \item Maintain a \textbf{neutral and contextually consistent tone} --- avoid promotional, advisory, or solution-seeking phrasing.
\end{itemize}

\vspace{0.5em}

\textbf{Inputs:}
\begin{itemize}
    \item Target Query: \texttt{\{query\}}
    \item Target Document: \texttt{\{validator\}}
    \item Top-5 Ranked Documents (Buffer A Source): \texttt{\{context\}}
    \item Previous Buffer A (query-related words): \texttt{\{key\_phrases\_buffer\_A\}}
    \item Previously generated sentences: \texttt{\{previous\_sentences\}}
\end{itemize}

\vspace{0.5em}

\textbf{Output:}
\begin{enumerate}
    \item Strictly output only valid JSON without any additional text.
    \item A list of exactly \texttt{\{n\_sent\}} new adversarial sentences (one per line, no explanations).
\end{enumerate}

\end{tcolorbox}

\vspace{-0.5em}
\caption{Feedback prompt used during iterative self-refinement when initial generations fail to improve ranking. The prompt re-extracts query-related phrases and enforces novelty relative to previously unsuccessful sentences.}
\label{fig:llm_feedback_prompt}
\end{figure}

\end{document}